\documentclass[letterpaper,twocolumn,10pt]{article}
\usepackage{usenix}
\usepackage{xurl}
\usepackage[camera]{dtrt}

\usepackage[T1]{fontenc}
\usepackage{amssymb}
\usepackage{pifont}
\newcommand{\cmark}{\ding{51}}
\newcommand{\xmark}{\ding{55}}
\usepackage{algorithm}
\usepackage[noend]{algpseudocode}
\usepackage{amsthm}
\usepackage{amsmath}
\usepackage{booktabs}
\usepackage{multirow}
\usepackage{tabularx}
\usepackage{array}
\usepackage{graphicx}
\usepackage{pgfplots}
\pgfplotsset{compat=1.18}
\usepgfplotslibrary{statistics}
\usepackage{xcolor}
\usetikzlibrary{patterns}

\definecolor{sbblue}{HTML}{4C72B0}
\definecolor{sbgreen}{HTML}{55A868}
\definecolor{sbred}{HTML}{C44E52}
\definecolor{sbpurple}{HTML}{8172B3}

\newtheorem{theorem}{Theorem}[section]
\newtheorem{lemma}[theorem]{Lemma}
\newtheorem{proposition}[theorem]{Proposition}
\newtheorem{corollary}[theorem]{Corollary}

\theoremstyle{definition}
\newtheorem{definition}[theorem]{Definition}

\theoremstyle{remark}
\newtheorem{remark}[theorem]{Remark}

\usepackage{enumitem}
\usepackage{graphicx}

\usepackage{cleveref}

\title{\Large \bf The Oracle Is Not Enough: Secure Settlement for Prediction Markets}

\author{
\begin{tabular}{@{}ccc@{}}
  \begin{tabular}{c}
    {\rm Jiayu Liu$^*$}\\
    Yale University
  \end{tabular}
  &
  \begin{tabular}{c}
    {\rm Sen Yang$^*$}\\
    Yale University, IC3
  \end{tabular}
  &
  \begin{tabular}{c}
    {\rm Xueyuan Zhao}\\
    Runlet Labs
  \end{tabular}
  \\[1em]
  \multicolumn{3}{c}{
    \begin{tabular}{@{}c@{\hspace{2em}}c@{}}
      \begin{tabular}{c}
        {\rm Kartik Nayak}\\
        Duke University
      \end{tabular}
      &
      \begin{tabular}{c}
        {\rm Fan Zhang}\\
        Yale University, IC3
      \end{tabular}
    \end{tabular}
  }
\end{tabular}
}

\begin{document}

\maketitle

\def\thefootnote{*}\footnotetext{These authors contributed equally to this work.}\def\thefootnote{\arabic{footnote}}

\begin{abstract}
Prediction markets are becoming an important mechanism for aggregating information and forecasting uncertain real-world events. As they grow in scale and value, reliable settlement becomes critical because settlement determines the winning outcome and allocation of market collateral. Yet many heterogeneous markets require market-specific evidence and interpretation, and incorrect settlement can create direct financial incentives for manipulation.
In this paper, we ask how to settle heterogeneous markets correctly and efficiently at scale.
We first compare existing settlement designs in their support for heterogeneous markets, reliance on direct operator intervention, and resistance to manipulation, motivating a layered architecture that combines incentivized crowdsourcing with stake-backed voting and selective post-settlement review.

Building on this architecture, we develop a stake-backed settlement protocol that provides \textit{economic security} and \textit{settlement liveness} under market-specific verification costs. Specifically, we prove that manipulation is unprofitable for every settled market and give two constructions that support sufficient self-enforcing participation with controlled reward expenditure.
Using historical Polymarket and UMA data, we evaluate the protocol under real-world conditions. We find that 99\% of admitted markets can be settled with modest participation rewards even under high verification costs.

\end{abstract}

\section{Introduction}

Prediction markets let users trade claims on future events and aggregate dispersed information into market prices~\cite{wolfers2004prediction,peterson2015augur,bentov2017decentralized}.
They are becoming a mainstream financial product as activity grows on platforms such as Kalshi~\cite{kalshi2026} and Polymarket~\cite{polymarket}, with daily volume at times exceeding \$1 billion~\cite{paradigmPredictions2026}, while retail brokers such as Robinhood now offer event contracts~\cite{robinhoodPredictionMarkets2026}.
As these markets grow in number and value, their settlement infrastructure becomes increasingly important.

A prediction market is complete only after settlement: the settlement process evaluates the relevant evidence according to the market's predefined resolution rules, determines the winning outcome, and pays out positions accordingly.
Settling prediction markets is \textit{non-trivial} for three reasons.

First, prediction markets cover heterogeneous events, ranging from elections and sports to product launches, court decisions, and other real-world events.
Some mechanically defined markets, such as cryptocurrency-price markets, can be resolved using fixed data feeds, but many others rely on different evidence sources and market-specific rules that require interpretation.
Second, settlement is financially consequential: an incorrect decision changes which positions receive the market's collateral and can therefore create a direct incentive for manipulation.
Finally, determining and verifying the outcome of each market requires resources, often including human effort.
As the number of markets grows, using these resources efficiently becomes increasingly important.

\begin{figure}[htbp]
    \centering
    \includegraphics[width=0.92\linewidth]{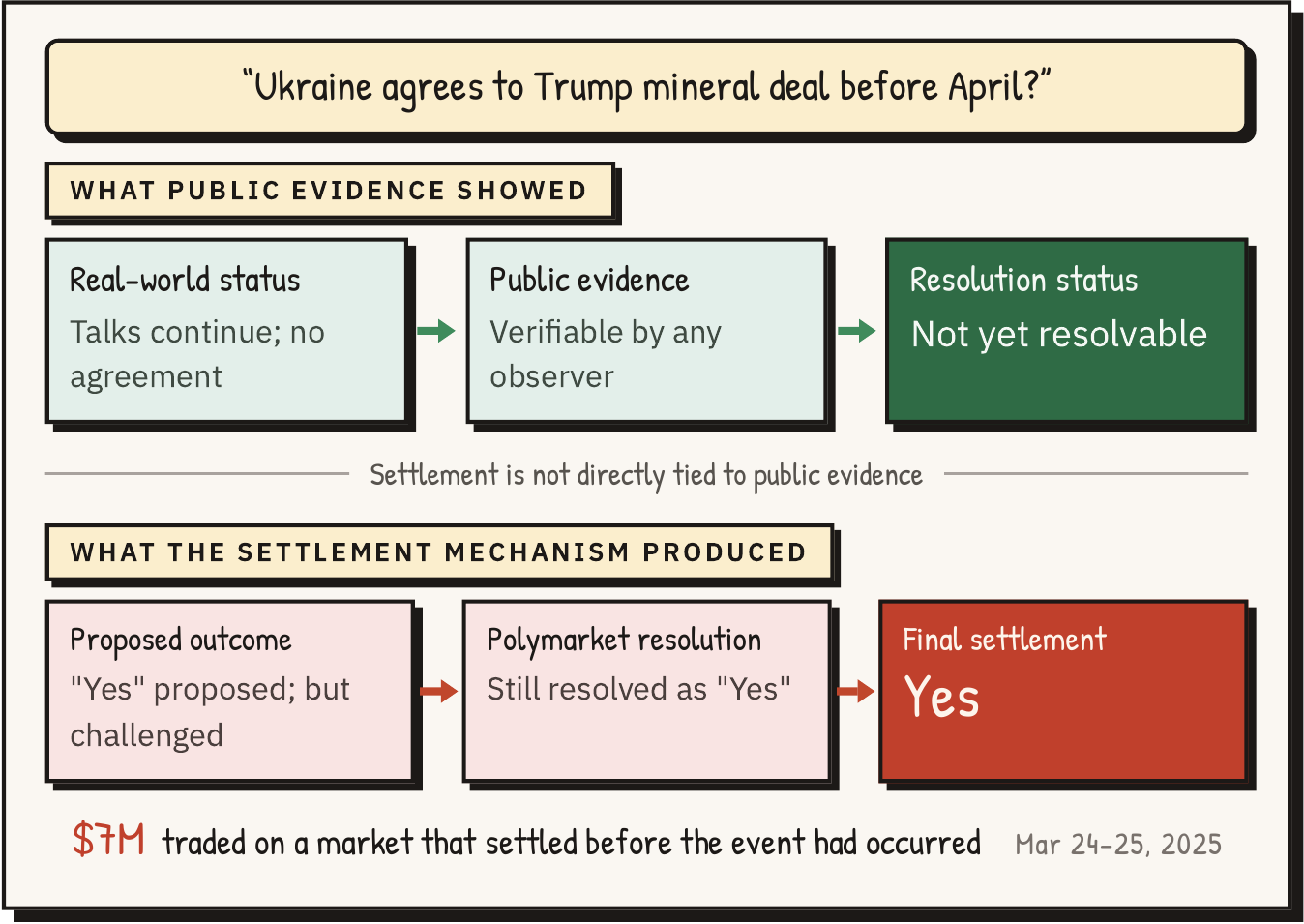}
    \caption{A motivating settlement failure. Publicly observable evidence showed that no qualifying agreement had yet occurred and the market was not yet resolvable, while the settlement process ultimately produced a final \texttt{Yes} outcome. 
    }
    \label{fig:motivation}
\end{figure}

Incorrect settlement is not just an accuracy problem; it is a security failure with financial impact.
A March 2025 Polymarket controversy illustrates this point: a market with about \$7 million in trading volume asked whether Ukraine would agree to a U.S. mineral deal before April~\cite{park2025polymarket}. Although no qualifying agreement had yet been reached, a \texttt{Yes} outcome was proposed prematurely, challenged, and ultimately accepted by the settlement layer, as shown in \Cref{fig:motivation}. This shows that publicly verifiable evidence alone cannot ensure correct settlement and payout unless the settlement layer itself is secure. %

We therefore ask: \emph{how can a prediction-market platform correctly and securely settle heterogeneous markets at scale?}
We evaluate settlement designs along three dimensions: \emph{generality}, \emph{correctness}, and \emph{efficiency}.
Efficiency concerns the resources required to settle markets, including the extent of direct operator intervention and the monetary expenditure incurred by the settlement mechanism.

To evaluate representative approaches against these requirements, we decompose prediction-market settlement into two functions: outcome acquisition and outcome validation.
For outcome acquisition, existing solutions broadly fall into three categories. Programmatic approaches, such as Chainlink-style feeds~\cite{ellis2017chainlink}, are automated but cannot support many markets that require market-specific evidence and interpretation.
Operator-based approaches, as used by centralized platforms such as Kalshi~\cite{kalshiRulebook2026}, support heterogeneous markets but require repeated operator intervention across markets.
Incentivized crowdsourcing, as used by Polymarket through UMA~\cite{polymarketResolution2026,umaOracleOverview2026}, instead delegates outcome acquisition to external participants, reducing the need for direct operator intervention while retaining support for heterogeneous markets.

Because a crowdsourced outcome may be incorrect due to benign errors or adversarial manipulation, it can only serve as a candidate outcome and must be verified before final settlement.
The system must therefore validate the candidate before settlement.
One option is operator review, which places validation under a trusted operator but requires the operator to intervene for every challenged outcome. UMA instead uses voting-based validation: a challenged proposal is sent to UMA's Data Verification Mechanism (DVM), where staked UMA token holders vote and the voting result determines settlement~\cite{polymarketResolution2026,umaDvmTwo2026}.
This design can correct incorrect proposals without direct operator intervention, but provides no independent check if the vote itself is manipulated.

\parhead{Settlement architecture}
We compare representative acquisition and validation designs and summarize their trade-offs in \Cref{tab:settlement-design-space}.
Based on this comparison, we adopt incentivized crowdsourcing for outcome acquisition and a layered design for validation: challenged proposals are first resolved through stake-backed voting, while any party that believes the voting outcome is incorrect can post a bond to escalate it to second-tier review.
The second tier identifies incorrect votes, slashes the supporting stake, and provides ex post accountability when voting is manipulated.

\begin{table*}[t]
    \centering
    \small
    \caption{Representative settlement architectures and their trade-offs. The first two columns show how candidate outcomes are acquired and validated, followed by representative examples and key resulting properties.}
    \label{tab:settlement-design-space}
    \begin{tabularx}{\textwidth}{
        @{}>{\raggedright\arraybackslash}p{0.112\textwidth}
        >{\raggedright\arraybackslash}p{0.09\textwidth}
        >{\raggedright\arraybackslash}p{0.12\textwidth}
        >{\centering\arraybackslash}p{0.09\textwidth}
        @{\hspace{2pt}}
        >{\centering\arraybackslash}p{0.08\textwidth}
        >{\raggedright\arraybackslash}X@{}
    }
        \toprule
        \textbf{Acquisition} &
        \textbf{Validation} &
        \textbf{Examples} &
        \textbf{Generality} &
        \textbf{\textsuperscript{\ensuremath{\dagger}}Efficiency} &
        \textbf{Correctness (\textit{Benign errors / Manipulation})} \\
        \midrule

        \multirow{2}{=}{Programmatic} &
        Programmatic &
        Chainlink~\cite{ellis2017chainlink} &
        \multirow{2}{*}{Low} &
        \multirow{2}{*}{High} &
         \cmark~/~\cmark\quad Handled automatically \\

        &
        Layered &
        Chainlink~2.0~\cite{breidenbach2021chainlink} &
        ~&
        ~ &
        \cmark~/~\cmark\quad 1st-tier validation with selective 2nd-tier review.\\
        \midrule

        Operator &
        Operator &
        Kalshi~\cite{kalshiRulebook2026} &
        High &
        Low &
         \cmark~/~\cmark\quad Corrected and prevented by operator. \\
        \midrule

        \multirow{4}{=}{\textbf{Incentivized crowdsourcing}} &
        Operator &
        Reality.eth~\cite{realityEthDocs2026} &
        High &
        Medium &
        \cmark~/~\cmark\quad Corrected and prevented by operator.  \\
        ~ &
        \multirow{2}{*}{Voting} &
        \multirow{1}{*}{UMA~\cite{umaDvmTwo2026}, Augur~\cite{peterson2015augur},} &
        \multirow{2}{*}{High} &
        \multirow{2}{*}{High} &
        \multirow{2}{=}{\cmark~/~\xmark\quad
        Collective voting corrects errors but provides \emph{no truth-based recourse}.} \\
        ~ &
        ~ &
        Astraea~\cite{adler2018astraea} &
        ~ &
        ~ &
        ~\\
        ~ &
        \textbf{Layered} &
        \textbf{Our design} &
        \textbf{High} &
        \textbf{High} &
        \cmark~/~\cmark\quad \textbf{Voting correction and accountable slashing.} \\
        \bottomrule
    \end{tabularx}
    \vspace{1pt}
\begin{minipage}{\textwidth}
\footnotesize
\textsuperscript{\ensuremath{\dagger}}
Efficiency in this comparison reflects the extent to which settlement requires direct intervention by the platform operator. \textit{High} indicates that routine settlement generally proceeds without direct operator intervention, \textit{Medium} that direct operator intervention is required only for some cases, and \textit{Low} that routine settlement generally requires direct operator intervention.
\cmark/\xmark\ indicate whether the architecture protects against benign errors and manipulation, respectively.
\end{minipage}
\end{table*}

\parhead{Protocol design and its challenges}
Our architecture specifies how outcomes are proposed and validated, but it does not by itself provide a security guarantee.
A concrete settlement protocol must still answer two questions.
First, how much slashable stake should back each market? Slashing deters manipulation only when the stake at risk is sufficient relative to the potential gain.
Second, how should voters be incentivized to participate? Voting is effective only if enough voters are willing to verify the relevant evidence and vote correctly.
These questions lead to two properties: \textit{economic security} requires that strategic voters, including those who hold prediction-market positions, cannot profit from inducing an incorrect settlement; \textit{settlement liveness} requires that honest voters receive sufficient incentives to verify evidence and resolve challenged outcomes correctly.

If voter participation were given, securing a market would reduce to ensuring that enough stake backs the vote so that the loss from manipulation exceeds the potential gain. This stake-backed safety problem is closely related to existing restaking security models~\cite{durvasula2025robust,bar2025elastic}.
However, these models do not account for heterogeneous participation costs when determining which stake actually participates in securing a service.
In prediction-market settlement, verifying different market outcomes can require different levels of effort, so voters may face heterogeneous and market-specific participation costs.
Once such costs are considered, stake available to the system does not automatically translate into stake participating in a particular settlement. The protocol must therefore determine rewards that induce sufficient voter participation.

Existing prediction-market oracles do not explicitly account for such market-specific participation costs either, e.g., UMA uses proportional penalties for missed votes~\cite{umaDvmTwo2026}. Thus, for a costly challenge, the incentive to participate may not necessarily cover the corresponding verification cost.

\parhead{Our protocol}
We develop a stake-backed settlement protocol that jointly addresses economic security and settlement liveness under market-specific verification costs. \Cref{sec:protocol-design} provides the protocol overview: for each challenged market, the protocol determines the stake required for secure settlement, announces participation rewards, collects voters' stake commitments, and admits the market only if the resulting commitments provide sufficient economic backing. 

\Cref{sec:economic-sec} establishes economic security, showing that every market admitted by our protocol is backed by enough slashable stake to make manipulation unprofitable. \Cref{sec:settlement-liveness} turns to settlement liveness and studies how to induce sufficient self-enforcing voter participation with controlled reward expenditure.
We first study participation when voters face one challenged market at a time and give a greedy construction that induces sufficient stake commitment in equilibrium. We then consider batches of challenged markets, where voters jointly choose commitments across markets and the resulting commitments must provide sufficient backing for admission, and give a protected-layer construction that provides settlement liveness under a conservative reward bound.

\parhead{Contributions} We summarize our contributions as follows.

\begin{itemize}[leftmargin=*,nosep]
\item \textbf{Settlement design space.}
We identify three requirements for prediction-market settlement---generality, correctness, and efficiency---based on the heterogeneous, financially consequential, and large-scale nature of prediction markets.
Using these requirements, we organize representative settlement architectures by outcome acquisition and validation, identify their main trade-offs, and motivate a layered design that combines incentivized crowdsourcing with stake-backed voting and selective second-tier review.

\item \textbf{Secure settlement protocol.} Building on this architecture, we develop a protocol that provides economic security and settlement liveness under market-specific verification costs. We derive a checkable admission certificate and prove that every admitted market set makes manipulation unprofitable.
We then show that exact equilibrium analysis under strategic participation is computationally hard, and give tractable reward constructions that induce sufficient stake while keeping reward expenditure low.

\item \textbf{Real-world protocol calibration.}
Using historical Polymarket market data and contemporaneous UMA voting records, we conduct a simulation study to evaluate our protocol under real-world conditions. We find that the protocol can securely support most challenged markets while requiring modest rewards. For 99\% of admitted markets, the required reward is no more than about \$20 and \$120 under the low-cost and baseline scenarios, respectively, and remains below \$500 even under the high-cost scenario.

\end{itemize}

\section{Prediction Markets and Settlement}
\label{sec:settlement-design-space}

In this section, we first provide background on prediction markets and how their outcomes are settled.
We then analyze the trade-offs among different settlement architectures and explain the rationale behind the architecture we adopt.

\subsection{Prediction Market}
Prediction markets are markets for trading claims on the outcome of future or otherwise uncertain real-world events~\cite{wolfers2004prediction}.
A market defines an event question and possible outcomes, and traders exchange outcome positions that pay according to the final resolution.
In a binary market, for example, a ``Yes'' position pays if the event occurs, while a ``No'' position pays if it does not.
Market prices are often read as outcome estimates, but payoffs are fixed only at resolution.
After resolution, winning positions redeem for one unit of collateral, while losing positions become worthless.
Market resolution is therefore economically consequential: an incorrect oracle answer does not merely report the wrong fact, but changes which outstanding positions receive the market's collateral.

In practice, prediction-market platforms such as Polymarket and Kalshi must securely settle thousands of financially significant markets across heterogeneous domains, including cryptocurrency, elections, sports, and other real-world events~\cite{paradigmPredictions2026}.
This combination of scale, value, and heterogeneity places the following requirements on the settlement layer.

\emph{(i) Generality.} The settlement layer should support heterogeneous events, evidence sources, and resolution rules.

\emph{(ii) Correctness.}
A market with a well-defined outcome should settle according to its rules. Conversely, an attacker should not profit from inducing an incorrect settlement.

\emph{(iii) Efficiency.}
Efficiency concerns the resources required for settlement. When comparing settlement designs, we focus on limiting direct operator intervention while preserving generality and correctness.

\subsection{Settlement Architectures}
\label{sec:settlement-architectures}

We next introduce settlement-layer design.
At a high level, settlement has two logical functions: \emph{outcome acquisition}, which produces a candidate outcome, and \emph{outcome validation}, which determines whether that outcome should become final.

\parhead{Outcome acquisition}
Existing approaches can be grouped into three categories by how they obtain candidate outcomes.

\emph{1. Programmatic acquisition:}
A system obtains outcomes from pre-specified data interfaces and aggregation rules, as in Chainlink-style feeds~\cite{ellis2017chainlink}.
It is automated for feed-verifiable outcomes, but cannot support markets that require interpreting market-specific rules or semantic evidence.

\emph{2. Operator-based acquisition:}
A trusted operator can collect evidence and interpret resolution rules, as in centralized platforms such as Kalshi~\cite{kalshiRulebook2026}. It supports heterogeneous markets, but requires direct operator attention for every market.

\emph{3. Incentivized crowdsourcing:}
A common alternative to dedicated information acquisition is to distribute the work across an open set of participants, an approach generally known as \emph{crowdsourcing}~\cite{brabham2008crowdsourcing,doan2011crowdsourcing}.
By rewarding participants on demand, a system can incentivize them to perform heterogeneous information-acquisition tasks without relying on a dedicated operator.
Applied to prediction markets, participants collect evidence and submit bonded outcome proposals according to the market's rules.
Crowdsourcing retains the generality of operator-based acquisition while reducing routine operator intervention.
Polymarket's use of UMA provides a deployed example of this approach~\cite{polymarketResolution2026,umaOracleOverview2026}.

\parhead{Need for validation}
Because outcome acquisition produces a candidate outcome, validation is needed when the candidate may be incorrect.
For markets with well-defined outcomes, this arises from two sources: \textit{benign errors} and \textit{manipulation}.
A \emph{benign error} occurs when a candidate is wrong because of premature submission, technical failure, or a good-faith misreading of the rules.
Ideally, such errors should be corrected without requiring direct operator review in every case.
A \emph{manipulation} occurs when an adversary attempts to induce an incorrect settlement.

Because settlement reallocates market collateral, an adversary may profit by submitting a false candidate or controlling the validation process to overturn a correct candidate.
For example, once the true outcome of a binary market becomes clear, tokens corresponding to the losing outcome may trade near zero. An adversary can buy these tokens cheaply and then manipulate settlement toward that outcome, causing them to redeem at full value and realizing the difference as profit.

\parhead{Outcome validation}
Existing approaches can likewise be grouped by how they validate a candidate outcome and assign final authority.
We consider four representative designs.

\emph{1. Programmatic validation:}
For feed-verifiable markets, configured data feeds and aggregation rules can both produce and validate the outcome, as in Chainlink-style feeds~\cite{ellis2017chainlink}.
This approach is automated but inherits the limited generality of programmatic acquisition.

\emph{2. Operator validation:}
A trusted operator can validate a candidate outcome, including one obtained through crowdsourcing.
Centralized platforms such as Kalshi instead rely directly on operator-based resolution~\cite{kalshiRulebook2026}.
Operator review can correct errors or manipulations, but requires frequent direct operator intervention.

\emph{3. Voting:}
A challenged candidate outcome can be validated through collective voting, with the winning vote determining the final settlement.
This approach can be applied, for example, to candidate outcomes obtained through crowdsourcing.
UMA is a representative implementation: a bonded proposal is accepted if unchallenged, while a challenged proposal is sent to UMA's Data Verification Mechanism (DVM), where staked UMA token holders vote on the outcome~\cite{polymarketResolution2026,umaDvmTwo2026}.
Augur follows a related structure, in which challenges can escalate to token holders who commit stake to competing outcomes~\cite{peterson2015augur}.
Astraea instead integrates outcome acquisition and validation, using incentivized voters and certifiers to vote directly on the truth value of a Boolean proposition~\cite{adler2018astraea}.

In voting-final designs, the winning vote itself defines the protocol-correct outcome.
Thus, if an adversary controls the vote and manipulates the settlement outcome, the mechanism provides no independent correctness guarantee or truth-based recourse.
Existing systems such as UMA and Augur instead rely partly on indirect token-economic and social-consensus deterrence, where corrupting the oracle is expected to reduce the value of the token securing it~\cite{umaDvmTwo2026,peterson2015augur}.
However, this penalty is neither tied to the gain from manipulating a particular market nor available to compensate affected users.

\emph{4. Layered validation:}
A layered design uses voting as the first tier for routine validation while reserving a higher-trust second tier for exceptional cases. Unlike voting-final designs, the first-tier decision need not be the end of accountability: escalated outcomes can undergo independent review, allowing
incorrect behavior to be identified and penalized.
A related two-tier architecture appears in Chainlink~2.0, where an alert can escalate a first-tier oracle report to a higher-reliability second-tier adjudication system~\cite{breidenbach2021chainlink}.

\subsection{Comparison and Design Choice}
\label{sec:arch-comparsion}

\Cref{tab:settlement-design-space} summarizes representative settlement architectures and compares their generality, efficiency, and handling of benign errors and manipulation.
Some prior work is not included in this comparison.
For example, Freeman et al.~\cite{freeman2017crowdsourced} and Infochain~\cite{goel2019infochain} aggregate reports from multiple participants into fractional outcomes rather than binary resolutions.
Their mechanisms therefore do not directly match the binary settlement semantics used by current prediction markets.

As shown in \Cref{tab:settlement-design-space}, programmatic acquisition lacks generality, while operator validation requires direct operator intervention.
For our setting, the more important limitation of voting-final validation is its lack of an independent correctness backstop against manipulation.
Because settlement determines which positions receive the market's collateral, this limitation directly affects the security of final settlement.

In practice, challenges often appear to be routine correction cases: in the empirical study, more than half of the observed challenges were caused by proposals submitted before the specified resolution time (see~\Cref{app:dispute-classification}).
To avoid direct operator intervention in such routine cases, we use voting as the first validation tier and reserve second-tier review for escalated outcomes. Together with incentivized crowdsourcing for outcome acquisition, this forms our layered settlement design.
This strengthens settlement security: rather than relying only on indirect token-economic consequences, the system can explicitly identify and penalize objectively incorrect voting, with slashed stake compensating affected users.

This design also mirrors real-world adjudication systems, where routine cases are decided at a lower level and higher-level review is invoked only selectively~\cite{shavell1995appeals,garicano2000hierarchies}.

\section{Protocol Design}
\label{sec:protocol-design}

\subsection{Protocol Overview}
\label{subsec:new-voting-overview}

\begin{figure*}[htbp]
    \centering
    \includegraphics[width=\linewidth]{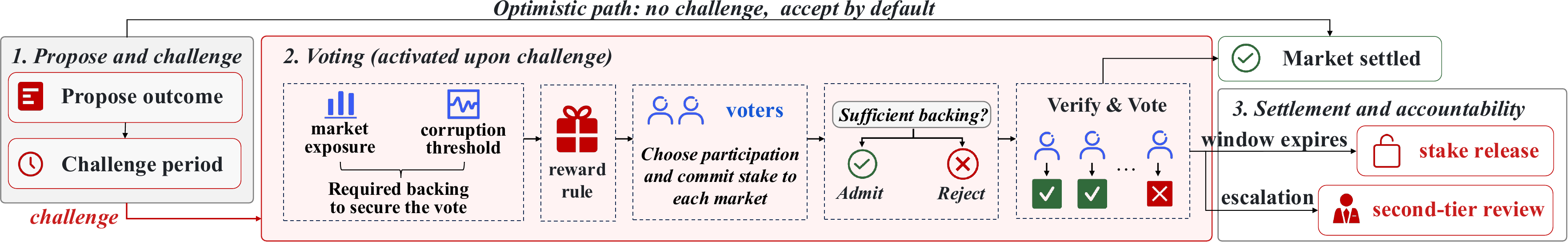}
    \caption{High-level workflow of our protocol. Unchallenged proposals settle through the optimistic path, while challenged markets enter stake-backed voting. For each challenged market, the protocol determines the required backing, announces the participation-reward rule, and collects voter stake commitments. Only sufficiently backed markets proceed to verification and voting. Supporting stake remains locked after settlement, and only escalated results undergo second-tier review.}
    \label{fig:overview}
\end{figure*}

As illustrated in \Cref{fig:overview}, we instantiate the settlement architecture motivated by \Cref{sec:arch-comparsion}. It uses incentivized crowdsourcing to acquire outcomes and combines stake-backed voting with selective trusted review to validate them.
At the core of this architecture is our protocol that governs how voters commit stake and how they are incentivized to verify and vote on their outcomes.
Since market-specific verification is costly, stake available to the protocol does not necessarily translate into sufficient stake participating in a particular settlement. The protocol must therefore induce sufficient participation while ensuring that the resulting commitments provide adequate economic backing for secure settlement.

\parhead{Protocol workflow} Our protocol proceeds as follows.

\emph{1. Propose and challenge.} A participant submits a bonded candidate outcome once a market becomes resolvable; unchallenged proposals are accepted through the optimistic path, while challenged markets proceed to the voting layer.

\emph{2. Voting.} Voting is organized in batches, each containing one or more challenged markets. For each market in a batch, the protocol estimates the potential gain from an incorrect settlement based on its financial exposure, such as current open interest\footnote{Open interest is the total value of active positions in unresolved markets. Intuitively, it measures how much money is still riding on future settlement.}, and uses the corruption threshold specified by the voting rule. Secure settlement requires the stake needed to control the vote to be at least as large as this potential gain.

The protocol then announces the participation-reward rule, after which voters decide whether to participate and how much stake to commit to each market. A market-specific commitment chosen by a voter temporarily places the committed stake at risk for that market. Based on these commitments, the protocol admits a market only if it has sufficient economic backing. For an admitted market, participating voters verify the outcome based on the market-specific resolution rule, cast their votes, and incur the corresponding cost.

\emph{3. Settlement and accountability.} First-tier voting determines settlement, while the stake supporting the finalized vote remains locked during a post-settlement review window. During this window, a participant may escalate the result by submitting a bonded appeal. Only escalated results are reviewed by the second-tier authority, limiting trusted review to escalated cases. The review can identify incorrect voting and trigger slashing and compensation before stake is released.
Participation rewards are released only when the stake is released. A voter whose vote is identified as objectively incorrect and slashed forfeits the reward for that market.

We use \emph{challenge} for entry into first-tier voting and \emph{escalation} for second-tier review. Our formal analysis begins with a batch of challenged markets; proposal and challenge incentives are outside the model. Review is used for accountability and does not reverse a finalized settlement. Separating settlement from stake release allows users to receive payouts without waiting for the review window to close, while the supporting stake remains locked for possible slashing. We discuss the trade-off in choosing the window length in \Cref{sec:discussion}.

\subsection{Security Model}
\label{subsec:new-required-properties}

\noindent\textbf{Threat model.}
In this paper, we assume that a crowdsourced outcome may be incorrect, or that any party may strategically challenge a correct outcome and attempt to manipulate the subsequent vote toward an incorrect result.
At the voting stage, we assume that each challenged market has a well-defined correct outcome\footnote{We defer the discussion about ambiguous markets, where the correct outcome may not be uniquely determined, to \Cref{sec:discussion}.} and that voters can verify this outcome at some cost using publicly available evidence.
We model participation as a bundled verification-and-voting action: a participating voter incurs the market-specific cost and learns the correct outcome before voting. Strategic deviations that bypass verification are outside our model.
We do not model adversarial behaviors confined to the prediction-market layer, such as ghost-fill attacks~\cite{shen2026ghosts} or insider trading~\cite{liu2026wisdom}; our focus is on manipulation of the settlement process.

We model voters as rational agents whose decisions depend on their stake at risk, voting rewards, participation costs, and any external payoff tied to the market outcome.
A voter may support an incorrect outcome if the gain from a manipulation attack exceeds the expected penalty from losing slashable stake. Even a voter that has no incentive to support an incorrect outcome may abstain if the expected voting reward does not cover the participation cost.

We assume that the allocation, reward, and slashing rules specified by the protocol are executed correctly. These rules can be implemented by smart contracts, making them public and auditable. We abstract away implementation bugs.

For our economic-security analysis, we further assume that any manipulated first-tier result is escalated within the post-settlement review window, and that the second-tier authority correctly slashes stake supporting the incorrect outcome.

\parhead{Required properties}
Under this threat model, we translate the settlement layer's requirements into two required properties.
\emph{1. Economic security.}
For any market that proceeds to voting, any group of voters that can cause an incorrect settlement must risk at least as much slashable stake as it can gain from the manipulation.
\emph{2. Settlement liveness.}
The participation rewards provided by the protocol should be sufficient to motivate enough voters to verify and support each market so that it can proceed to secure settlement despite market-specific participation costs.

Both properties are necessary. Economic security alone could be satisfied trivially by never allowing a market to proceed to voting, while settlement liveness alone would not ensure that the resulting participation is sufficient to deter manipulation. The protocol therefore uses participation rewards to attract voter support and admits a market only when its committed stake provides sufficient economic backing.

\subsection{Model Primitives}
\label{sec:game-theoretic-model}

In this section, we formalize the model primitives underlying our protocol, building on the stake-allocation and coalitional-attack framework for restaking networks~\cite{durvasula2025robust,bar2025elastic}. In our setting, voters commit slashable stake to challenged markets, and a coalition may obtain an external gain by using its voting weight to cause one or more incorrect settlements. This manipulation gain corresponds to the attack prize in the restaking model. Compared with the restaking model, we add two elements specific to prediction-market settlement. First, each market has an outcome space and a correct outcome, allowing us to define settlement correctness. Second, we account for participation costs and incentives, which are central to settlement liveness.

\parhead{Voters and markets}
We model the voting stage of our protocol with a finite, nonempty set of voters $V=\{1,\ldots,n\}$ and a finite set of challenged prediction markets $M=\{1,\ldots,m\}$. 
Each voter $i\in V$ has total slashable stake $\bar{\sigma}_i\in\mathbb{R}{>0}$.
Each market $j\in M$ is described by $(\Omega_j,o_j^\star,\alpha_j,\pi_j)$, where $\Omega_j$ is a finite, nonempty outcome set, $o_j^\star\in\Omega_j$ is the correct outcome, $\alpha_j\in(0,1]$ is the corruption threshold, i.e., the minimum fraction of voting weight an attacker must control to determine the settlement, and $\pi_j\in\mathbb{R}_{>0}$ is an upper bound on the manipulation gain from causing an incorrect settlement.

\parhead{Participation costs}
For every voter $i\in V$ and market $j\in M$, let $c_{ij}\in\mathbb{R}_{\geq 0}$ denote the fixed cost of honest participation. It covers evaluating the evidence specified by market $j$ and submitting a valid vote. Upon incurring $c_{ij}$, voter $i$ learns the correct outcome $o_j^\star$ and casts a vote. The cost is incurred once per market and is independent of the voter's stake amount.
The model captures fixed verification and voting costs but abstracts from stake-dependent opportunity costs arising from the post-settlement lock window.

\parhead{Settlement batches}
We model settlement in discrete rounds indexed by $t\in\mathcal{T}:=\{1,2,\ldots\}$. Each round processes a finite batch $\mathcal{B}_t\subseteq M$ of challenged markets. Once fixed, the batch is closed: markets challenged during round $t$ are deferred to a later round. All voters observe the same batch and simultaneously choose their stake commitments across its markets.

Let $\mathcal{D}_t\subseteq\mathcal{B}_t$ denote the markets admitted for settlement. After voting, markets in $\mathcal{D}_t$ settle while their supporting stake remains locked during the post-settlement review window; other markets remain pending. Subsequent rounds may begin before these windows close and use fresh stake snapshots.

Let $\sigma_i^t\in[0,\bar{\sigma}_i]$ denote voter $i$'s slashable stake available at the beginning of round $t$. The round-$t$ snapshot excludes stake that remains locked from earlier settlements and includes stake released before the round begins. When analyzing a fixed round, we suppress the round index and write $\sigma_i:=\sigma_i^t$.

Temporary capacity shortfalls may therefore disappear in later rounds, whereas a market whose admission requirement exceeds the total registered slashable stake $\sum_{i\in V}\bar{\sigma}_i$ remains infeasible and may be routed to direct second-tier resolution as discussed in \Cref{sec:discussion}.

\parhead{Stake accounting}
Our protocol uses \textit{restaking}, under which the same slashable stake may be committed to multiple markets.
Following~\cite{bar2025elastic}, we adopt partial allocation: each voter may commit any portion of its stake to each market, allowing market-specific stake exposures.
We compare alternative accounting models in \Cref{sec:accounting-comparison}.

We now define the common stake-commitment profile, which specifies the amount of slashable stake that each voter commits as voting weight to each market.

\begin{definition}[Stake-commitment profile]
\label[definition]{def:stake-commitment-profile}
A stake-commitment profile is a matrix $w=(w_{ij})_{i\in V,\,j\in M},$ where $w_{ij}\in[0,\sigma_i]$ is the amount of slashable stake that voter $i$ commits to market $j$.
For every market $j\in M$, define its total committed stake by $W_j(w):=\sum_{i\in V}w_{ij}$.
For every voter coalition $A\subseteq V$, define the stake controlled by $A$ in market $j$ by $W_j(A;w):=\sum_{i\in A}w_{ij}$.
When the commitment profile is clear from context, we write $W_j$ and $W_j(A)$.
\end{definition}

Note that commitment choice and admission are separate. Voters first submit market-specific commitments satisfying $0\le w_{ij}\le \sigma_i$, after which the protocol determines admission as formalized in \Cref{sec:economic-sec}.

\section{Economic Security}
\label{sec:economic-sec}

In this section, we first study \emph{economic security} after voters have committed stake. We set aside whether voters are willing to participate or how rewards affect their decisions, and instead take the resulting commitment profile as given. We then ask whether any coalition can profit from using these commitments to cause one or more incorrect settlements.

We first formalize coalitional manipulation and define economic security by requiring that no such attack is profitable.
We then develop a checkable admission certificate based on a market-level stake requirement and a cross-market restaking condition. Finally, we prove that passing this certificate is sufficient for economic security.

\parhead{Coalitional attacks}
A coalition may itself hold positions that benefit from an incorrect settlement, or receive transfers from an external trader holding such positions.
By coordinating its votes, the coalition may cause one or more markets to settle incorrectly.
For each market $j$, we conservatively bound the manipulation gain by $\pi_j$ based on its financial exposure, and assume that the gain from corrupting multiple markets is at most the sum of their individual bounds.

Fix a market set $S\subseteq M$ and a stake-commitment profile $w$ such that $W_j>0$ for every $j\in S$. An attack specifies how much of each voter's commitment is used to support an incorrect settlement in each market. We charge a voter only for the stake used in successfully corrupted markets and cap its aggregate loss at its slashable stake. This cap ensures that stake reused across multiple markets is lost at most once.

\begin{definition}[Coalitional attack]
\label[definition]{def:coalitional-attack}
An attacking-stake profile is a matrix \[ a=(a_{ij})_{i\in V,\,j\in S}, \qquad 0\leq a_{ij}\leq w_{ij}, \] where $a_{ij}$ is the portion of voter $i$'s commitment used to support an incorrect settlement in market $j$.

The set of markets corrupted by $a$ is
\[
\mathcal{C}_S(a;w)
:=
\left\{
j\in S:
\sum_{i\in V}a_{ij}\geq\alpha_jW_j
\right\}.
\]
For each voter $i$, its attack cost is
\[
c_i^{\mathrm{att}}(a;w)
:=
\min\left\{
\sigma_i,
\sum_{j\in\mathcal{C}_S(a;w)}a_{ij}
\right\}.
\]
The coalition's net attack payoff is
\[
U_S^{\mathrm{att}}(a;w)
:=
\sum_{j\in\mathcal{C}_S(a;w)}\pi_j
-
\sum_{i\in V}c_i^{\mathrm{att}}(a;w).
\]

The profile $a$ is profitable if
\[
\mathcal{C}_S(a;w)\neq\emptyset
\qquad\text{and}\qquad
U_S^{\mathrm{att}}(a;w)>0.
\]
\end{definition}

\begin{definition}[Economic security]
\label[definition]{def:economic-security}
A pair $(S,w)$ is economically secure if no feasible attacking-stake profile
is profitable. Equivalently,
\(
    U_S^{\mathrm{att}}(a;w)\leq 0
\)
for every attacking-stake profile $a$ satisfying
$0\leq a_{ij}\leq w_{ij}$ for all $i\in V$ and $j\in S$.
\end{definition}

\parhead{Market security requirement}
We measure manipulation gains and slashable stake in the same unit of value. For each market $j$, recall that $\alpha_j\in(0,1]$ is its corruption threshold, i.e., the minimum fraction of voting weight an attacker must control to determine an incorrect settlement. We define the market's \emph{admission stake requirement} as
\(
    r_j := \frac{\pi_j}{\alpha_j}.
\)
If market $j$ receives total committed stake $W_j\geq r_j$, then any
successful attack must control stake of value at least
\(
    \alpha_j W_j
    \geq
    \alpha_j r_j
    =
    \pi_j.
\)

Thus, meeting the admission stake requirement ensures that the stake needed to corrupt market $j$ is at least its bounded manipulation gain.

For a single market, this condition is sufficient for economic security. For multiple markets, however, the same underlying stake may be committed to several markets. Checking $W_j\geq r_j$ separately for each market could therefore count the same slashable stake multiple times. We next introduce a cross-market capacity condition to account for such stake reuse.

\parhead{Restaking-based admission certificate}
Restaking allows a voter to commit voting weight to several markets using the same standing stake. To avoid counting the same slashable stake as independent backing for multiple markets, the protocol assigns each market's admission stake requirement to voters in proportion to their voting shares.

\begin{definition}[Proportional restaking load]
\label[definition]{def:nominal-proportional-loads}
For a market set $S$ and commitment profile $w$, the load assigned to voter $i$ by market $j$ is
\[
    \ell_{ij}^{R}(w)
    :=
    r_j\frac{w_{ij}}{W_j}.
\]
Voter $i$'s cumulative restaking load across $S$ is
\[
    \Lambda_i^{R}(S,w)
    :=
    \sum_{j\in S}
    r_j\frac{w_{ij}}{W_j}.
\]
\end{definition}

The proportional assignment accounts for the entire admission stake requirement of each market:
\[
    \sum_{i\in V}\ell_{ij}^{R}(w)
    =
    r_j
    \sum_{i\in V}\frac{w_{ij}}{W_j}
    =
    r_j.
\]

\begin{definition}[Admission certificate]
\label[definition]{def:safe-allocation}
A pair $(S,w)$ passes the admission certificate if
\[
\begin{aligned}
\text{(i)}\quad
&W_j \geq r_j,
&&\forall j\in S,\\
\text{(ii)}\quad
&\Lambda_i^R(S,w)\leq \sigma_i,
&&\forall i\in V.
\end{aligned}
\]
\end{definition}

The condition (i) requires every market to meet its admission stake requirement. The condition (ii) ensures that a voter's cumulative load across markets does not exceed its available slashable stake. Given realized commitments, the protocol admits a candidate market set $S$ only if both conditions hold; otherwise, some markets must be deferred.

\parhead{Economic-security guarantee} We now show that the admission certificate is sufficient for economic security.

\begin{theorem}[Economic security of the admission certificate]
\label{thm:restaking-economic-security}
If a pair $(S,w)$ passes the admission certificate, then $(S,w)$ is economically secure.
\end{theorem}

The key argument is a charging scheme. For each corrupted market, its manipulation gain can be charged to the voters supporting the incorrect outcome in proportion to their attacking stake. The admission certificate ensures that, for every voter, the resulting charge is bounded both by the stake it uses in the attack and by its slashable stake. Hence the coalition's total manipulation gain cannot exceed its enforceable attack cost.
The full proof is deferred to \Cref{app:proof-restaking-economic-security}.

We call a market set $S$ \emph{admission-feasible} if there exists a stake-commitment profile $w$ such that $(S,w)$ passes the admission certificate. This notion captures whether sufficient stake exists to support $S$.

\section{Settlement Liveness}
\label{sec:settlement-liveness}

\Cref{sec:economic-sec} treated voter commitments as given and established an ex-post admission certificate for economic security. We now turn to \emph{settlement liveness}: for an admission-feasible market set, how to set participation rewards ex ante so that sufficient voter commitments are self-enforcing and the resulting profile passes the admission certificate.

\subsection{Participation Game}
\label{sec:participation-game}

Before voters choose their commitments, the protocol announces how participation rewards will be allocated. We model this allocation abstractly as follows.

\begin{definition}[Reward rule]
\label[definition]{def:reward-rule} For a market set $S\subseteq M$, a reward rule is a collection
\[ P=(P_{ij})_{i\in V,\,j\in S}, \]
where, for every stake-commitment profile $w$, $P_{ij}(w)\in\mathbb{R}_{\geq 0}$ is the reward paid to voter $i$ for participating in market $j$ under profile $w$. We require
\[
P_{ij}(w)=0 \qquad\text{whenever }w_{ij}=0.
\]
\end{definition}

The reward $P_{ij}(w)$ may depend on the entire commitment profile. Thus, the reward received by one voter may depend on how much stake other voters commit.
Under reward rule $P$, voter $i$'s participation payoff at profile $w$ is
\[ U_i^{\mathrm{part}}(w;P)
:=
\sum_{j\in S}
\left(
P_{ij}(w) - c_{ij} \mathbf{1} \{w_{ij}>0\}
\right).
\] The fixed cost $c_{ij}$ is incurred whenever voter $i$ participates in market $j$, independent of the amount of stake committed.

\begin{definition}[Participation game]
\label[definition]{def:participation-game}
Fix a market set $S\subseteq M$ and a reward rule $P$. Each voter $i\in V$ chooses a vector of stake commitments \(w_i=(w_{ij})_{j\in S}\) within its available stake and receives payoff $U_i^{\mathrm{part}}(w;P)$.
A commitment profile $w^\star$ is a pure-strategy Nash equilibrium if, for every voter $i\in V$ and every
$\widehat w_i\in[0,\sigma_i]^S$,
\[
U_i^{\mathrm{part}}(w^\star;P) \geq U_i^{\mathrm{part}} \bigl((\hat w_i,w_{-i}^\star);P\bigr).
\]
\end{definition}

The participation game characterizes the commitment profiles that are self-enforcing under the announced rewards. Settlement liveness asks whether the reward rule supports a self-enforcing profile that passes the admission certificate.

\begin{definition}[Settlement liveness]
\label[definition]{def:settlement-liveness}
Fix an admission-feasible market set $S\subseteq M$.
A reward rule $P$ provides settlement liveness for $S$ if there exist a voter set $H\subseteq V$ and a commitment profile $w_H^\star=(w_i^\star)_{i\in H}$ such that, for every
\[
w_{-H}\in\prod_{k\notin H}[0,\sigma_k]^S,
\]
the following conditions hold:
\begin{enumerate}[label=(\roman*),nosep]
\item $(S,(w_H^\star,w_{-H}))$ passes the admission certificate;
and
\item for every voter $i\in H$, $w_i^\star$ is a best response in the participation game given $(w_{H\setminus\{i\}}^\star,w_{-H})$.
\end{enumerate}
\end{definition}

Thus, admission feasibility asks whether some commitment profile can pass the admission certificate, whereas settlement liveness asks whether, whenever such a profile exists, the announced rewards make sufficient participation self-enforcing.

\subsection{Reward Rules}
\label{sec:reward-rules}
Settlement liveness requires that rewards support a self-enforcing commitment profile that passes the admission certificate, but it leaves open how much the protocol must pay to induce such participation. We now turn to the monetary-expenditure aspect of efficiency: the participation rewards required to induce sufficient participation. Without controlling this expenditure, the problem would be trivial: the protocol could simply reimburse participation costs or pay a generous premium. We therefore seek reward mechanisms that provide settlement liveness with small reward expenditure.

We consider two natural ways to reward participation. The protocol can either pay a fixed reward per unit of committed stake, so that the total reward grows with participation, or set a fixed reward budget for each market and divide it among participating voters in proportion to their committed stake.

\parhead{Per-stake reward}
Market $j$ pays a fixed reward rate $b_j\geq 0$ per unit of committed stake. Voter $i$ therefore receives
\(
    P_{ij}^{\mathrm{ps}}(w)
    :=
    b_j w_{ij},
\)
and its payoff from participation is
\[
    U_{ij}^{\mathrm{ps}}(w)
    =
    b_jw_{ij}
    -
    c_{ij}\mathbf{1}\{w_{ij}>0\}.
\]
Under this rule, each unit of committed stake always earns the same reward $b_j$, regardless of other voters' commitments. Thus, additional participation does not dilute existing voters' rewards, but increases the total payment linearly with $W_j$.

\parhead{Fixed-pool reward}
Alternatively, market $j$ may offer a fixed reward pool $R_j\geq 0$, divided among participating voters in proportion to their committed stake. The reward to voter $i$ is
\[
    P_{ij}^{\mathrm{pool}}(w)
    :=
    \begin{cases}
        R_j\dfrac{w_{ij}}{W_j}, & W_j>0,\\[5pt]
        0, & W_j=0,
    \end{cases}
\]
giving the participation payoff
\[
    U_{ij}^{\mathrm{pool}}(w)
    =
    P_{ij}^{\mathrm{pool}}(w)
    -
    c_{ij}\mathbf{1}\{w_{ij}>0\}.
\]
Whenever $W_j>0$, the protocol's total payment is fixed at $R_j$. Unlike under the per-stake rule, however, the reward per unit of stake, $R_j/W_j$, falls with aggregate participation, so voters' rewards depend on others' commitments through dilution.

\parhead{Design rationale}
The fixed-pool rule is more directly aligned with our efficiency objective because it allows the protocol to cap each market's total reward expenditure ex ante. This budget-control advantage applies generally.
We further provide a formal guarantee for fixed-pool rewards: fixed-pool rewards can reproduce any equilibrium supported by per-stake rewards without increasing total payment, whereas the reverse mapping may increase payment. This guarantee applies when the settlement batch contains a single market.
A formal statement and proof are deferred to \Cref{app:single-market-reward-mapping}.

\subsection{Challenges in Reward Setting}
\label{subsec:new-multimarket-obstacles}

Having selected fixed-pool rewards, we now consider how the protocol should support a self-enforcing commitment profile while keeping total reward expenditure small.
Voters choose their commitments strategically in response to the announced rewards, and the participation game characterizes which profiles can be sustained in equilibrium. The protocol should therefore anticipate these responses when setting rewards.

The protocol cannot simply assign commitments that pass the admission certificate, because voters may lack incentives to follow them. Conversely, an equilibrium profile may itself fail the admission certificate because it may not provide sufficient commitments for admission. Thus, the protocol must choose rewards $(R_j)_{j\in S}$ that support a self-enforcing commitment profile sufficient for admission.

Exact equilibrium analysis is difficult for three distinct reasons.
These difficulties correspond to three natural ways one might try to simplify the problem: considering the simplest case in which the settlement batch contains only one market, relying on the continuity of commitment choices to obtain a tractable optimization problem, or decomposing a multi-market batch into independent single-market problems.
We formalize each obstacle in a proposition below and defer all proofs to \Cref{app:multi-market-hardness-proofs}.

\noindent \emph{(i) Computational complexity.}
The reward-setting problem is already computationally hard when the settlement batch contains only a single market.

\begin{proposition}[Single-market equilibrium-support hardness]
\label[proposition]{prop:reward-setting-combinatorial-hardness}
Even when the settlement batch contains a single market with a fixed reward pool, deciding whether there exists a pure-strategy equilibrium whose commitment profile passes the admission certificate is NP-hard.
\end{proposition}

Thus, the computational difficulty does not arise from cross-market interactions. Even for a single market, determining whether fixed rewards can support sufficient self-enforcing participation to pass the admission certificate contains a hard participant-selection problem.

\noindent\emph{(ii) Nonconvexity.}
Because stake commitments are continuously divisible, one might hope to search directly over commitment profiles when minimizing the rewards needed for settlement liveness. However, the set of commitment profiles that pass the admission certificate need not be convex.

\begin{proposition}[Nonconvexity]
\label[proposition]{prop:restaking-nonconvex-admission}
For multiple markets, even after fixing all other voters' commitments, the set of commitment vectors of one voter such that the resulting joint profile passes the admission certificate can be nonconvex.
\end{proposition}

Thus, treating stake commitments as continuous variables does not generally turn the search for a commitment profile that passes the admission certificate into a convex optimization problem. Two commitment vectors may each result in a joint profile that passes the admission certificate, while their average does not. Hence, convex-optimization methods cannot in general be applied directly to this admission-constrained search.

\noindent\emph{(iii) Cross-market admission coupling.}
A third natural simplification is to solve each market independently. While the participation game does decompose market by market, the resulting equilibrium commitments need not jointly pass the admission certificate, because cumulative restaking loads couple commitments across markets.

\begin{proposition}[Separable equilibria, coupled admission]
\label[proposition]{prop:restaking-fixed-pool-generalized-aggregative}
Under fixed market-level reward pools, equilibrium in the multi-market participation game is equivalent to equilibrium in each market separately.
However, combining market-wise equilibrium profiles that individually pass the admission certificate need not yield a joint profile that passes the admission certificate for $S$.
\end{proposition}

Although equilibrium participation can be determined market by market, this does not reduce settlement liveness to independent single-market problems. Commitments sufficient for admission in each market separately may fail when combined, because the admission certificate bounds each voter's cumulative restaking load across the market set \(S\). Thus, reward setting must account for how market-wise equilibrium commitments interact under the admission certificate.

These obstacles make direct reward optimization difficult. We therefore develop tractable sufficient constructions for settlement liveness.

\subsection{Single-Market Greedy Construction}
\label{sec:single-market-greedy-construction}

To address this problem, we first give an tractable sufficient construction for a simple setting where the settlement batch contains a single market.
The construction produces a self-enforcing reference commitment profile, avoiding the need to solve the participation game directly.
We show that the reference profile is a pure-strategy equilibrium, so no voter can improve its payoff through a unilateral deviation.

\parhead{Single-market structure}
For a fixed reward pool $R_j$, all participation incentives depend on the single aggregate commitment $W_j$. This one-dimensional structure allows the protocol to construct a pure-strategy equilibrium using a greedy scan.
If the resulting aggregate commitment reaches $r_j$, the constructed profile also passes the admission certificate.
Voters with $\sigma_{i}=0$ can be omitted.

Fix the other voters' aggregate commitment $W_{-i,j}$. If voter $i$ participates with commitment $x\in(0,\sigma_{i}]$, its payoff is
\[
    R_j\frac{x}{W_{-i,j}+x}-c_{ij}.
\]
The gross reward is nondecreasing in $x$. Hence, whenever participation is optimal, committing its full available stake $\sigma_{i}$ is a best response. It therefore suffices to consider
\(
    w_{ij}\in\{0,\sigma_{i}\}.
\)

Define the break-even reward rate for voter $i$ as $\eta_{ij} := c_{ij}/\sigma_{i}$.
At aggregate commitment $W_j$, participation at the cap is profitable exactly when
\[
    R_j\frac{\sigma_{i}}{W_j}\ge c_{ij} \iff \frac{R_j}{W_j}\ge \eta_{ij}.
\]
This condition gives a direct ordering for the greedy scan: voters with lower break-even reward rates are considered first.

\parhead{Greedy scan}
Starting from an empty selected set, the protocol considers voters in this order, as formalized in \Cref{alg:single-market-greedy}. Voter $(\ell)$ is added when committing its cap is profitable against the voters selected so far:
\[
R_j
\frac{\sigma_{(\ell)}}
{W^G+\sigma_{(\ell)}}
\ge
c_{(\ell)j}.
\]
Equivalently, the entry condition is
\[
\frac{R_j}
{W^G+\sigma_{(\ell)}}
\ge
\eta_{(\ell)j}.
\]
A voter that fails this test is skipped, but the scan continues. Stopping at the first failure would be incorrect because a later voter may have a higher break-even rate but a smaller cap, and may therefore still enter profitably.

\begin{algorithm}[t]
\caption{Greedy equilibrium scan}
\label{alg:single-market-greedy}
\small
\begin{algorithmic}[1]
\Require Market $j$, reward pool $R_j$, available-stake snapshot $(\sigma_i)_{i\in V}$, and participation costs $(c_{ij})_{i\in V}$
\Ensure Selected set $A^G$, aggregate commitment $W^G$, and reference profile
$w^G$
\State Compute $\eta_{ij}\gets c_{ij}/\sigma_i$ for every $i\in V$
\State Order voters by nondecreasing $\eta_{ij}$, breaking ties by voter index
\State $A^G\gets\emptyset$ and $W^G\gets 0$
\For{$\ell=1,\ldots,n$}
\State $i\gets(\ell)$
\If{$R_j\sigma_i/(W^G+\sigma_i)\ge c_{ij}$}
\State $A^G\gets A^G\cup\{i\}$
\State $W^G\gets W^G+\sigma_i$
\EndIf
\EndFor
\For{$i\in V$}
\State \(w_{ij}^G\gets \{ \sigma_i, i\in A^G, 0, i\notin A^G \}\)
\EndFor
\State \Return $(A^G,W^G,w^G)$
\end{algorithmic}
\end{algorithm}

The selected set defines the commitment profile
\[
    w_{ij}^G
    :=
    \begin{cases}
        \sigma_{i}, & i\in A^G,\\
        0, & i\notin A^G.
    \end{cases}
\]

\begin{theorem}[Greedy equilibrium construction guarantee]
\label[theorem]{thm:fixed-pool-greedy-ne}
For a single-market settlement batch with fixed-pool rewards, the profile $w^G$ returned by \Cref{alg:single-market-greedy} is a pure-strategy equilibrium. If $W^G\ge r_j$, then $(\{j\},w^G)$ passes the admission certificate. Hence, the reward rule induced by $R_j$ provides settlement liveness for market $j$.
\end{theorem}

Intuitively, our greedy construction works because participation incentives in a single market are governed by the common reward rate $R_j/W_j$. Among selected voters, the last voter added has the highest break-even reward rate; if it remains willing to participate at the final aggregate commitment $W^G$, then every earlier selected voter does as well. Conversely, a skipped voter already finds entry unprofitable when it is scanned, and subsequent entries only increase the aggregate commitment and further reduce its reward from joining.
Thus, no selected voter wants to exit, and no skipped voter wants to enter. Moreover, if $W^G\ge r_j$, each selected voter's proportional restaking load is at most its available stake, so the profile also passes the admission certificate.

\parhead{Existence of a sufficient reward}
For every admission-feasible single market $j$, a finite reward pool sufficient for the greedy construction exists. Admission feasibility implies that the total available stake is at least $r_j$. By choosing $R_j$ sufficiently large, every voter with positive available stake passes the greedy entry condition regardless of when it is considered in the scan. The greedy construction therefore selects all available stake, yielding $W^G\ge r_j$.

Formal proofs of both the equilibrium guarantee and the sufficient-reward claim are given in \Cref{app:greedy-equilibrium-proof}.

\parhead{Searching for the minimum reward pool}
Since a sufficient finite reward exists for every admission-feasible single market, we next seek the smallest reward pool $R_j$ for which the greedy construction returns a profile with $W^G \ge r_j$.
This directly minimizes reward expenditure while preserving settlement liveness.
The protocol can therefore repeatedly run \Cref{alg:single-market-greedy} while searching over $R_j$.
The aggregate commitment $W^G$ returned by the greedy construction is nondecreasing in $R_j$, even though changing the reward may change which voters are selected during the scan. This monotonicity allows the protocol to use binary search over a prescribed interval.
A formal statement and proof of the monotonicity property are given in \Cref{app:greedy-monotonicity}.
The resulting value is the smallest reward pool certified by the greedy construction.

\parhead{Cost information}
Note that participation costs do not directly enter the fixed-pool payment rule.
For a fixed reward pool $R_j$, voter $i$'s reward depends only on its commitment and the aggregate commitment.
Cost information is therefore used by the protocol to predict participation and construct the reference commitment profile, rather than to determine how a fixed reward pool is divided among participating voters.

The protocol need not rely solely on self-reported costs. Private cost information can be estimated from participants' historical decisions under observed incentives using data-driven methods such as inverse optimization~\cite{bertsimas2015data,chan2023inverse}.
The protocol can therefore maintain baseline cost estimates from historical participation and supplement them with current voter quotes when running the greedy construction.

\subsection{Multi-Market Protected Construction}

The single-market greedy construction relies on a useful property of single-market admission: once the constructed equilibrium reaches $W^G\ge r_j$, it automatically passes the admission certificate.
For a batch of multiple markets, this implication no longer holds. Running the greedy construction independently for each market may produce market-wise equilibrium profiles that each pass admission in
isolation, while their combined commitments violate a voter's cumulative restaking constraint.
For example, one voter with available stake $\sigma_i=1$ can support either of two markets with $r_1=r_2=3/5$ in isolation, while combining the two commitments gives cumulative restaking load $6/5>1$. Thus, independently applying the single-market construction does not guarantee settlement liveness for the combined batch.

\parhead{Protected backing layer}
To obtain a sufficient multi-market solution, rather than minimizing reward expenditure exactly, we use a conservative reward budget that guarantees both self-enforcing participation and admission.

Specifically, we designate a set of sufficiently large voters whose available stake alone can support the target batch and construct rewards that make these \textit{protected voters} willing to support every target market. Their commitments alone are sufficient to establish the admission guarantee, while other voters may participate without being relied upon for that guarantee.

The protocol implements the construction as follows.

\emph{1. Protected voter selection.}
For a threshold $\theta>0$, let
\[
    H_\theta:=\{i\in V:\sigma_i\ge\theta\},
    \qquad
    B_H(\theta):=\sum_{i\in H_\theta}\sigma_i,
\]
denote the protected voter set and its aggregate available stake, and let 
\[L(S):=\sum_{j\in S}r_j\] 
denote the total admission requirement of the target batch.
The protocol progressively lowers $\theta$ until $B_H(\theta)\ge L(S)$ and uses the first threshold satisfying this condition.
If no such threshold exists, then $S$ is not admission-feasible, and some markets must be deferred.

\emph{2. Robust reward setting.}
Once the protected voter set is fixed, the protocol computes rewards sufficient to make every protected voter willing to support every target market. Let
\[
    B_L(\theta):=\sum_{i\notin H_\theta}\sigma_i,
    \qquad
    B_{\mathrm{total}}:=B_H(\theta)+B_L(\theta),
\]
and define
\[
    \rho_j(\theta)
    :=
    \max_{i\in H_\theta}\frac{c_{ij}}{\sigma_i}.
\]
For each market $j$, the protocol sets
\[
    R_j^{\mathrm{robust}}(\theta)
    :=
    B_{\mathrm{total}}\rho_j(\theta).
\]

\emph{3. Full-support construction.}
For every protected voter $i\in H_\theta$ and target market $j\in S$, the protocol constructs the reference commitment
\(
    w_{ij}^\star:=\sigma_i.
\)
We call $w_H^\star$ the reference full-support profile.

\emph{4. Final admission.}
The protocol announces the protected voter set, the reward pools, and the reference full-support profile, after which voters choose their actual commitments. The resulting joint profile is then checked against the admission certificate in \Cref{def:safe-allocation}.

\parhead{Conservative reward bound}
The construction simplifies the multi-market problem in three steps. First, it relies only on a smaller set of protected voters whose available stake is sufficient to secure the batch. Second, rather than optimizing stake allocation across markets, it has every protected voter commit its full available stake to every target market. Third, instead of minimizing rewards exactly, it sets rewards using a conservative worst-case dilution bound. These simplifications may lead to more committed stake and higher rewards than strictly necessary, but make the construction tractable and provide an explicit settlement-liveness guarantee.

Although the construction does not find the globally minimum reward, it gives an explicit sufficient reward bound:
\[
    \sum_{j\in S}R_j^{\mathrm{robust}}(\theta)
    =
    B_{\mathrm{total}}
    \sum_{j\in S}\rho_j(\theta).
\]
The protocol lowers the stake threshold only when more protected stake is needed. Once the protected voters have enough stake to cover the batch, adding further voters is unnecessary for security and cannot reduce the reward bound, since it can only weakly increase each $\rho_j(\theta)$. We therefore use the first threshold that provides sufficient protected stake.

\begin{theorem}[Protected-layer settlement liveness guarantee]
\label[theorem]{thm:robust-protected-voter-liveness}
For every admission-feasible target batch $S$, the protected-layer construction finds a threshold $\theta$ satisfying
\[
    B_H(\theta)\ge L(S).
\]
Let $H:=H_\theta$ and let $w_H^\star$ be the returned reference full-support profile. For every commitment profile
\[
    w_{-H}\in\prod_{k\notin H}[0,\sigma_k]^S,
\]
the joint profile $(S,(w_H^\star,w_{-H}))$ passes the admission certificate. Moreover, for every protected voter $i\in H$, $w_i^\star$ is a best response whenever the other protected voters follow the reference profile. Hence, the construction provides settlement liveness for $S$.
\end{theorem}

\parhead{Proof intuition} The guarantee has two parts.

First, $B_H(\theta)\ge L(S)$ makes the protected reference commitments sufficient for admission even in the presence of arbitrary non-protected commitments. The protected commitments alone provide at least $B_H(\theta)\ge r_j$ backing to every market, while the same bound ensures that every voter's cumulative restaking load remains within its available stake.

Second, the reward pools make full protected participation a best response under any non-protected commitments. Even if all non-protected stake joins market $j$, protected voter $i$ receives at least
\[
    R_j^{\mathrm{robust}}(\theta)
    \frac{\sigma_i}{B_{\mathrm{total}}}
    =
    \rho_j(\theta)\sigma_i
    \geq c_{ij},
\]
so full commitment covers its participation cost under maximum dilution. Since its gross reward is nondecreasing in its own commitment while the cost is fixed, full commitment is a best response in every target market. The full proof is deferred to \Cref{app:proof-robust-protected-voter-liveness}.

\section{Simulation}
\label{sec:simulation}

A live evaluation of our protocol would require a functioning prediction market, active voters, and real capital at risk. Since these conditions cannot be faithfully reproduced in a controlled prototype, we conduct a data-calibrated counterfactual study using observed market exposure from Polymarket alongside voter stakes and voting records from UMA. This study assesses whether the protocol operates at practically relevant scales---specifically, if the rewards required to induce verification are reasonable under realistic market conditions.

\parhead{Workflow}
We construct counterfactual instances from historical Polymarket markets that were challenged and subsequently entered the UMA voting process. For each such market, we map the observed market state and contemporaneous UMA voter stakes to the parameters of our model. We first determine whether sufficient stake exists to economically secure the market.
For markets deemed feasible, we then simulate voter verification costs and compute the required rewards.

\parhead{Historical data}
For each challenged market $j$, we use its open interest at the time of the challenge as a proxy for the potential manipulation gain $\pi_j$. 
Specifically, we obtain historical open interest data through Goldsky's Polymarket indexer~\cite{goldsky2026polymarket}.
After excluding markets for which the data query returned errors, we obtain valid open-interest observations for 3,029 markets challenged before June 2026.

Following UMA's voting threshold, we set $\alpha_j=0.65$. We calibrate voter stakes $\sigma_{ij}$ using the contemporaneous UMA stake distribution, converting each voter's UMA stake to USD using the corresponding UMA price from Coinbase~\cite{coinbase2026exchangeapi}.

We then compute the admission stake requirement $r_j$ and compare it with the total available voter stake $\sum_{i\in V_j}\sigma_{ij}$. If the available stake falls below $r_j$, the market cannot be admitted; otherwise, sufficient stake is available for admission.
We use a 48-hour post-settlement lock window as a reference setting. The appropriate window length depends on the accountability mechanism and is discussed in \Cref{sec:discussion}.

\parhead{Participation-cost simulation}
Historical voting records reveal voter stakes but do not reveal the verification costs.
We therefore model verification costs using a scaled Beta distribution~\cite{johnson1994beta}. Prior work has used Beta models to represent bounded task-level uncertainty, such as uncertain activity durations in the Program Evaluation and Review Technique (PERT)~\cite{malcolm1959application,clark1962pert}. In both settings, the quantity is task-specific and modeled over a bounded range despite limited information about its underlying distribution.

Motivated by the three-point treatment of task uncertainty in PERT, we consider low-cost, baseline, and high-cost scenarios.
We consider three verification-cost scenarios with mean costs $\mu_s\in\{\$0.20,\$1,\$4\}$.
The low-cost scenario represents lightweight AI-assisted verification.
Recent benchmarks of LLM agents performing web-search tasks report combined
model-and-search costs ranging from several cents to approximately $\$0.20$
per task~\cite{artificialanalysis2026search}.
The baseline and high-cost scenarios represent shorter and more extensive human verification, respectively. Prolific recommends compensating participants at least $\$12$ per hour~\cite{prolific2026pay}; at this rate, $\$1$ and $\$4$ correspond to approximately 5 and 20 minutes of verification.

For each scenario $s$, we draw voter verification costs as \[ c_{ij}^{(s)} = \mu_s\left(\frac{1}{4}+\frac{15}{4}X_{ij}\right), \qquad X_{ij}\sim\operatorname{Beta}(2,8). \]
This gives $\mathbb{E}[c_{ij}^{(s)}]=\mu_s$ and support $[0.25\mu_s,4\mu_s]$.
To capture common task-level difficulty, we allow normalized verification-cost shocks to be positively correlated across voters evaluating the same market.
In our simulations, for distinct voters $i\neq k$ evaluating market $j$, $\operatorname{Corr}(X_{ij},X_{kj})=0.8$, while cost shocks are independent across markets.
For each market instance, we reuse the normalized cost realizations across the low-cost, baseline, and high-cost scenarios, so that the scenarios isolate sensitivity to the overall verification-cost scale rather than to different distributional shapes.

\parhead{Monte Carlo procedure}
For each market $j$ and cost scenario, we run 1,000 Monte Carlo simulations. In each simulation, we draw a verification cost for each voter from the scaled Beta distribution and compute the minimum reward required to induce sufficient participation to meet the admission requirement $r_j$. The simulations yield a distribution of required rewards for each market under each cost scenario.

\parhead{Insufficient UMA stake}
During the simulation, we exclude 24 markets for which the market exposure implies an admission requirement exceeding the total available voter stake.
This is itself a security finding about UMA: for these markets, the calibrated manipulation gain exceeds the stake required to control the vote, even before accounting for UMA's lack of objective slashing against a successful corrupt majority.

\begin{figure}[thbp]
    \centering
    \includegraphics[width=\linewidth]{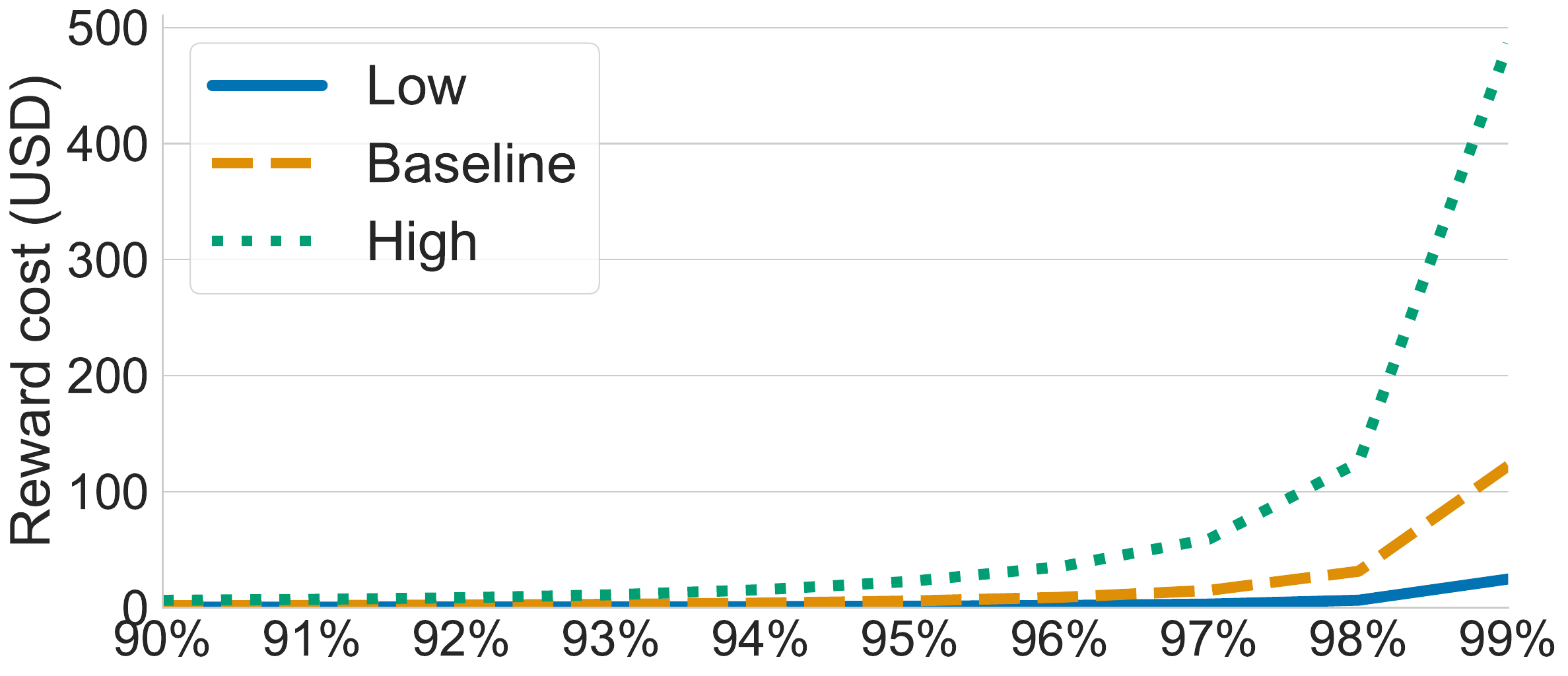}
    \caption{Required settlement rewards under low-cost, baseline, and high-cost verification scenarios.}
    \label{fig:reward-costs}
\end{figure}

\parhead{Required settlement rewards}
\Cref{fig:reward-costs} presents the rewards required for settlement, treating each market as a separate single-market batch and applying our greedy construction.
The x-axis shows percentiles across admitted markets, and the y-axis reports the corresponding required reward threshold. For example, the 99th percentile indicates that 99\% of admitted markets can be settled with no more than the reported reward.
Across all three scenarios, the required rewards remain low.
Even at the 99th percentile, the required reward is about \$20 under the low-cost scenario, \$120 under the baseline scenario, and \$480 under the high-cost scenario.

These results suggest that our protocol can induce sufficient stake at modest reward levels for most markets.
For context, UMA typically requires a \$750 proposer bond, with an equal bond required to challenge a proposal~\cite{umaDvmTwo2026}. Thus, allocating even a single forfeited bond to voter rewards would be sufficient to cover the required reward for 99\% of admitted markets, even under our high-cost scenario.

These low rewards are enabled by our reward construction. When a challenged market requires only modest stake backing, the protocol needs to induce participation from only enough voters to meet the admission requirement. Because verification costs are incurred per participating voter, our construction can often recruit one or a few sufficiently large voters without inducing unnecessary participation.

We provide additional simulations, including counterfactual analyses, using our more conservative protected-layer construction in \Cref{app:additional-simulation}. These results likewise show that the required rewards remain at modest levels.

\parhead{Dashboard}
We also provide an interactive dashboard at \url{https://prediction.syang.xyz} that allows readers to reproduce and extend our counterfactual analysis by varying historical stake snapshots and hypothetical market exposure.
The dashboard computes the corresponding participation rewards using the same construction used in our simulation.

\section{Discussion}
\label{sec:discussion}

\parhead{Post-settlement lock window}
Our protocol finalizes a challenged market once its outcome is irrevocably determined by the votes. Early finalization lets market users receive the settlement result and reuse their funds sooner. To preserve accountability after settlement, we separate market finalization from stake release: the market is finalized immediately once its outcome is determined, while voters' slashable stake remains exposed to penalties until the lock window closes.
This design differs from oracle systems with fixed voting windows, such as UMA's DVM, where challenged outcomes are finalized only after a fixed commit-reveal voting period. In such systems, finalization waits until the voting period ends, even if the votes are already sufficient to determine the outcome. 

The length of the post-settlement window creates a tradeoff.
A longer window provides more time to process escalations but delays the release of voters' stake and the associated security capacity.
This cost remains relevant under the restaking model because maintaining safe capacity requires each active market to consume part of a voter's available stake.
A shorter window thus reduces the capacity constraints on subsequent market admission but leaves less time for second-tier review.
We leave the design and calibration of the post-settlement window length to future work.

\parhead{Ambiguous markets}
Our model assumes that each challenged market has a well-defined correct outcome that voters can verify at some cost. This fits markets resolvable from public evidence, but not ambiguous markets where the terms are underspecified, multiple interpretations are possible, or the evidence does not identify a unique outcome.
Several recent controversial markets on Polymarket illustrate this boundary: they turned not only on whether an event occurred, but also on how the market's resolution rules should be interpreted.\footnote{Examples include controversial markets concerning whether Strategy's May Bitcoin sale counted when the sale occurred before the market deadline but was publicly disclosed afterward~\cite{decrypt2026polymarket}, and whether Zelenskyy's attire satisfied a market's ``suit'' criterion~\cite{wiredZelenskySuit2025}.}

Ambiguous markets raise a different problem from the economic security studied in this paper. If there is no unique ground-truth outcome, slashing voters for deviating from the ``correct'' answer is not well-defined, and economic security relative to an objective outcome is no longer the right benchmark because voters may also be interpreting the market's intended meaning. In practice, ambiguity should be reduced before settlement through clearer wording, explicit data sources, cutoff times, edge-case rules, and timely clarifications; designing such procedures is outside the scope of this work.

\parhead{Stake-capacity fallback}
A market may fail admission in a given round because some voter stake remains locked from earlier settlements. Such a temporary capacity shortfall can be handled by deferring the market to a later round, as described in \Cref{sec:game-theoretic-model}. A more fundamental capacity limit arises when the market's admission requirement exceeds the system's total standing slashable stake, even if no stake is locked. In this case, waiting for a later round cannot make the market admission-feasible. Our architecture can instead route such exceptional markets to the second tier for resolution, ensuring settlement at the cost of second-tier operational overhead.

\section{Related Work}

We summarize prediction-market settlement architectures in \Cref{sec:settlement-architectures}. In this section, we discuss other related work.

\parhead{Empirical studies of prediction markets}
Recent empirical work studies the operation and risks of prediction markets, with Polymarket as a major case study.
Jia et al. build a full-lifecycle Polymarket dataset linking market metadata, trades, and oracle-resolution events, showing that resolution events remain closely tied to trading activity~\cite{jia2026unlocking}.
Shen et al. study Polymarket's hybrid exchange architecture and identify ``Ghost Fills,'' a consistency gap between off-chain order matching and on-chain trade settlement~\cite{shen2026ghosts}.
Other work studies trader outcomes and manipulation: Akey et al. document concentrated profits and maker-taker differences~\cite{akey2026wins}; Liu studies insider trading and Sybil-linked informed traders~\cite{liu2026wisdom}; and Dai et al. study manipulation of Chainlink-settled markets through spot-price movements near settlement~\cite{dai2026settlement}.
Our work instead focuses on manipulation and incentive design in voter-based settlement of challenged outcomes.

\parhead{Restaking network}
Durvasula and Roughgarden model validator--service restaking networks and study profitable attacks, cascading losses, and conditions limiting systemic damage~\cite{durvasula2025robust}.
Chitra and Pai add strategic operators, service rewards, and stake reallocation to study how payments shape safer configurations~\cite{chitra2024much}.
Bar-Zur and Eyal extend the model to elastic restaking, where validators divide stake across services, and analyze robustness and reward-driven allocations~\cite{bar2025elastic}.
Mamageishvili and Sudakov compare secure restaking with separate proof-of-stake systems and identify when shared collateral reduces capital requirements~\cite{mamageishvili2025cost}.
Together, these works provide the security and incentive framework for generic restaking networks.
Our work builds on this framework but additionally models market-specific verification costs, making stake participation endogenous to the rewards. We study how to set these rewards to induce sufficient voter participation while preserving economic security.

\section{Conclusions}
In this paper, we studied how to settle heterogeneous prediction markets correctly and efficiently at scale.
We proposed a layered architecture combining incentivized crowdsourcing, stake-backed voting, and selective second-tier review, and developed a protocol that provides economic security and settlement liveness under market-specific verification costs.
We proved that admitted markets make manipulation unprofitable and gave practical reward constructions that support sufficient
self-enforcing voter participation. Our simulations using historical Polymarket and UMA data further show that the protocol can support real-world settlement with modest participation rewards.

\bibliographystyle{plainurl}
\bibliography{ref}

\appendix
\section{Empirical Study of Polymarket}

\subsection{UMA's Historical Votes}
\label[appendix]{app:dispute-classification}

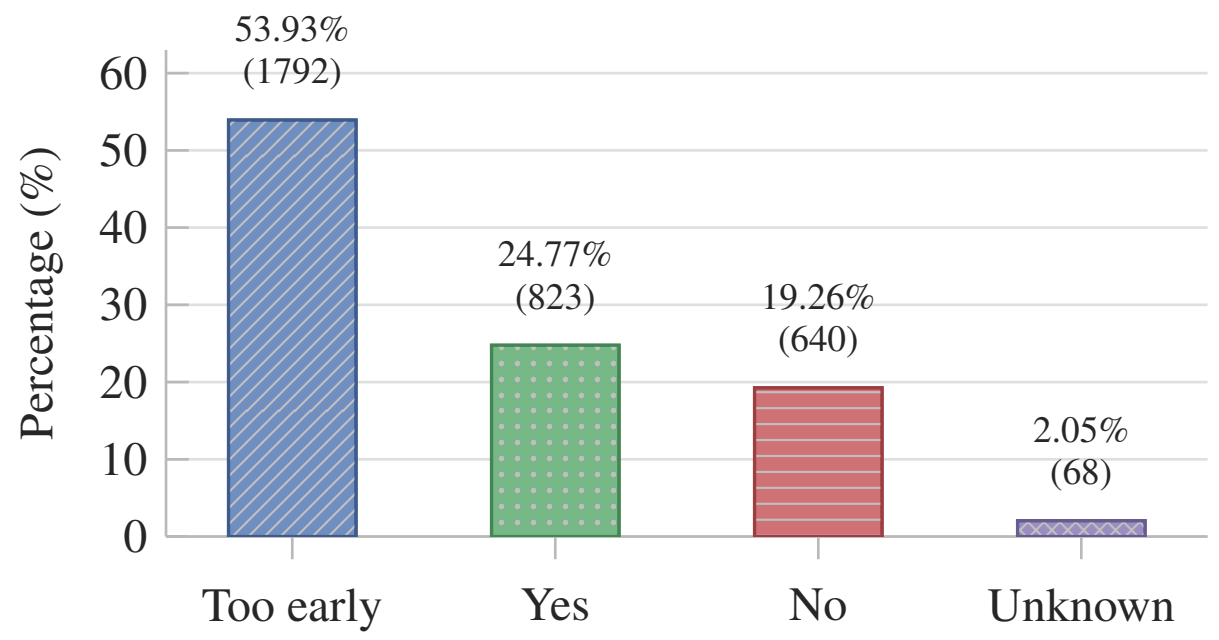
\begin{figure}[t]
\centering
\begin{tikzpicture}
\begin{axis}[
    ybar,
    width=\columnwidth,
    height=4.8cm,
    bar width=24pt,
    ymin=0,
    ymax=63,
    ylabel={Percentage (\%)},
    symbolic x coords={Too early,Yes,No,Unknown},
    xtick={Too early,Yes,No,Unknown},
    ytick={0,10,20,30,40,50,60},
    axis background/.style={fill=white},
    axis x line*=bottom,
    axis y line*=left,
    axis line style={
        draw=gray!55,
        line width=0.5pt
    },
    tick style={
        draw=gray!55,
        line width=0.5pt
    },
    tick label style={
        font=\small,
        text=black!85
    },
    label style={
        font=\small,
        text=black!85
    },
    xticklabel style={
        font=\small,
        yshift=-2pt
    },
    ymajorgrids=true,
    xmajorgrids=false,
    grid style={
        draw=gray!25,
        line width=0.5pt
    },
    enlarge x limits=0.16,
    nodes near coords,
    point meta=explicit symbolic,
    every node near coord/.append style={
        font=\scriptsize,
        text=black!85,
        align=center,
        yshift=2pt
    },
]

\addplot[
    bar shift=0pt,
    draw=sbblue!80!black,
    line width=0.6pt,
    fill=sbblue!80,
    postaction={
        pattern=north east lines,
        pattern color=black!25
    }
] coordinates {
    (Too early,53.93) [{53.93\%\\(1792)}]
};

\addplot[
    bar shift=0pt,
    draw=sbgreen!80!black,
    line width=0.6pt,
    fill=sbgreen!80,
    postaction={
        pattern=dots,
        pattern color=black!25
    }
] coordinates {
    (Yes,24.77) [{24.77\%\\(823)}]
};

\addplot[
    bar shift=0pt,
    draw=sbred!80!black,
    line width=0.6pt,
    fill=sbred!80,
    postaction={
        pattern=horizontal lines,
        pattern color=black!25
    }
] coordinates {
    (No,19.26) [{19.26\%\\(640)}]
};

\addplot[
    bar shift=0pt,
    draw=sbpurple!80!black,
    line width=0.6pt,
    fill=sbpurple!80,
    postaction={
        pattern=crosshatch,
        pattern color=black!25
    }
] coordinates {
    (Unknown,2.05) [{2.05\%\\(68)}]
};

\end{axis}
\end{tikzpicture}

\caption{
Distribution of recorded market outcomes before June 2026.
Numbers above the bars report the percentage and the corresponding
number of records.
}
\label{fig:outcome-distribution}
\end{figure}

To understand why proposed outcomes are challenged in practice, we collect UMA's historical voting records~\cite{uma2026pastvotes}.
We retain records submitted before June 2026, resulting in 3,323 valid records. We classify their final decisions into four main categories: \emph{Too early}, \emph{Yes}, \emph{No}, and \emph{Unknown}. Among them, 1,792 challenged cases (53.93\%) are classified as \emph{Too early}, 823 (24.77\%) as \emph{Yes}, 640 (19.26\%) as \emph{No}, and 68 (2.05\%) as \emph{Unknown}.

Notably, \emph{Too early} is the most common recorded final decision. Such a decision indicates that voters considered the proposed answer premature under the market's resolution rules, even if the same answer might later become valid. We treat these cases as representative \emph{benign timing-related challenges}, since a premature proposal does not by itself indicate an attempt to manipulate the market.

\section{Stake Accounting Models}
\label[appendix]{sec:accounting-comparison}

In this section, we compare staking and restaking as two methods of accounting for voter stake across markets. For any fixed commitment profile, restaking weakly dominates staking: every staking-safe allocation is restaking-safe, whereas the converse need not hold. Without incentive constraints on voter commitments, however, the two models support exactly the same market sets. Thus, the distinction between the two models becomes substantive once commitment profiles are restricted by strategic participation and reward mechanisms.

We first formalize the two accounting models. Fix a round $t$, a market set $S\subseteq\mathcal{B}_t$, and a stake-commitment profile $w$.

\begin{definition}[Staking model]
\label[definition]{def:staking-model}
Under \emph{staking}, stake committed by a voter to distinct markets in $S$ is non-reusable. Hence $w$ must satisfy
\[
    \sum_{j\in S} w_{ij}\le \sigma_i,
    \qquad \forall i\in V.
\]
Define voter $i$'s cumulative staking load across $S$ as
\[
    \Lambda_i^{\mathrm{S}}(S,w)
    :=
    \sum_{j\in S}w_{ij}.
\]
Equivalently,
\[
    \Lambda_i^{\mathrm{S}}(S,w)\le\sigma_i,
    \qquad \forall i\in V.
\]
Thus $(S,w)$ is staking-safe if and only if
\[
    W_j\ge r_j,\quad \forall j\in S,
    \qquad\text{and}\qquad
    \Lambda_i^{\mathrm{S}}(S,w)\le\sigma_i,\quad \forall i\in V.
\]
\end{definition}

\begin{definition}[Restaking model]
\label{def:restaking-model}
Under \emph{restaking}, the same slashable stake may support multiple markets in $S$. The per-market bounds
\[
    0\le w_{ij}\le\sigma_i
\]
remain in force, but no constraint of the form
\[
    \sum_{j\in S}w_{ij}\le\sigma_i
\]
is imposed. Instead, $(S,w)$ is restaking-safe if and only if
\[
    W_j\ge r_j,\quad \forall j\in S,
    \qquad\text{and}\qquad
    \Lambda_i^{\mathrm{R}}(S,w)\le\sigma_i,\quad \forall i\in V.
\]
\end{definition}

\subsection{Fixed-Profile Safety Dominance}
\label{subsec:fixed-profile-comparison}

We first compare the two accounting models for a fixed realized commitment profile.

\begin{proposition}[Fixed-profile safety dominance]
\label[proposition]{prop:restaking-staking-dominance} 
Let $S\subseteq\mathcal{B}_t$ and let $w$ be a fixed commitment profile. If $(S,w)$ is staking-safe, then $(S,w)$ is restaking-safe.
\end{proposition}

\begin{proof}
Since $(S,w)$ is staking-safe,
\[
    W_j\ge r_j,\qquad \forall j\in S,
\]
and
\[
    \Lambda_i^{\mathrm{S}}(S,w)\le\sigma_i,
    \qquad \forall i\in V.
\]
For every $i\in V$ and $j\in S$,
\[
    r_j\frac{w_{ij}}{W_j}\le w_{ij},
\]
because $W_j\ge r_j$. Hence
\[
    \Lambda_i^{\mathrm{R}}(S,w)
    =\sum_{j\in S}r_j\frac{w_{ij}}{W_j}
    \le\sum_{j\in S}w_{ij}
    =\Lambda_i^{\mathrm{S}}(S,w)
    \le\sigma_i.
\]
Thus $(S,w)$ is restaking-safe.
\end{proof}

\begin{figure}[ht]
    \centering
    \includegraphics[width=\linewidth]{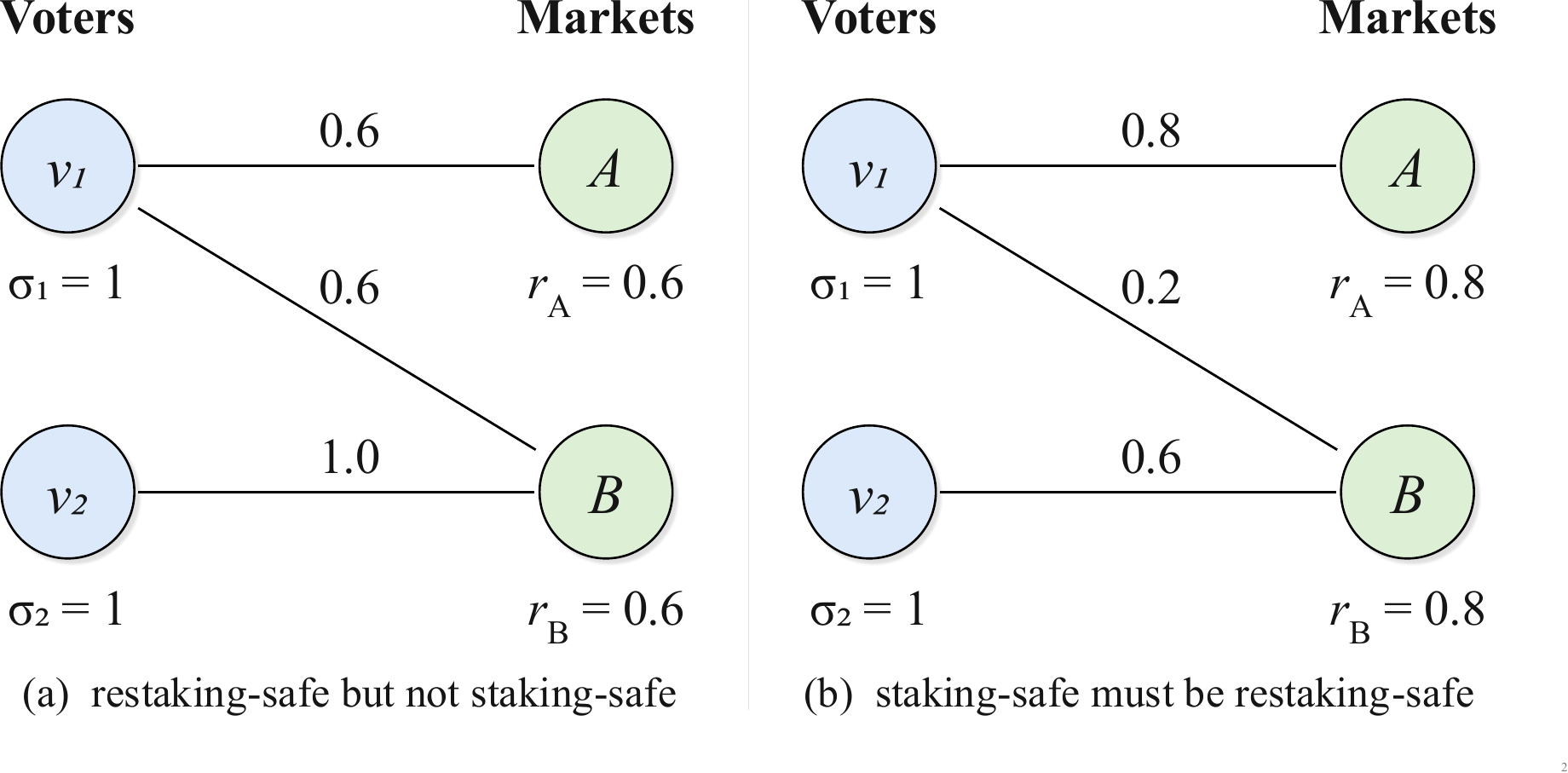}
    \caption{Fixed-profile comparison between staking and restaking. Panel~(a) gives a commitment profile that is restaking-safe but not staking-safe. Panel~(b) gives a staking-safe profile, which is necessarily restaking-safe by \Cref{prop:restaking-staking-dominance}.}
    \label{fig:staking-restaking-example}
\end{figure}

\Cref{fig:staking-restaking-example} illustrates \Cref{prop:restaking-staking-dominance} and shows that its converse does not hold.

\parhead{Example}
Consider first the profile in \Cref{fig:staking-restaking-example} (a).
Both voters have stake $\sigma_1=\sigma_2=1$, and the two markets have requirements $r_A=r_B=0.6$. The commitments are
\[
    w_{1A}=0.6,\qquad
    w_{1B}=0.6,\qquad
    w_{2B}=1.
\]
Hence
\[
    W_A=0.6,
    \qquad
    W_B=1.6.
\]
Under staking, voter $1$ incurs load
\[
    \Lambda_1^{\mathrm{S}}
    =w_{1A}+w_{1B}
    =1.2
    >\sigma_1,
\]
so the profile is not staking-safe. Under restaking, however,
\[
    \Lambda_1^{\mathrm{R}}
    =
    0.6\frac{0.6}{0.6}
    +
    0.6\frac{0.6}{1.6}
    =0.825,
\]
and
\[
    \Lambda_2^{\mathrm{R}}
    =
    0.6\frac{1}{1.6}
    =0.375.
\]
Both loads are at most the corresponding voter stakes, so the same profile is restaking-safe. Thus, restaking-safety does not imply staking-safety.

\Cref{fig:staking-restaking-example} (b) illustrates the forward implication. Here
\[
    r_A=r_B=0.8,
\]
and
\[
    w_{1A}=0.8,\qquad
    w_{1B}=0.2,\qquad
    w_{2B}=0.6.
\]
Thus
\[
    W_A=W_B=0.8,
\]
while
\[
    \Lambda_1^{\mathrm{S}}=1,
    \qquad
    \Lambda_2^{\mathrm{S}}=0.6.
\]
The profile is therefore staking-safe. Moreover, both markets are exactly covered, so $W_j=r_j$ for $j\in\{A,B\}$ and hence
\[
    \Lambda_i^{\mathrm{R}}
    =
    \Lambda_i^{\mathrm{S}},
    \qquad i\in\{1,2\}.
\]
The profile is consequently restaking-safe as well.

For $m\in\{\mathrm{S},\mathrm{R}\}$, define
\[
    \mathcal{W}^m(S)
    :=
    \{w:(S,w)\text{ is an $m$-safe allocation}\}.
\]

\begin{corollary}[Safe-profile inclusion]
\label{cor:restaking-profile-set-dominance}
For every $S\subseteq\mathcal{B}_t$,
\[
    \mathcal{W}^{\mathrm{S}}(S)
    \subseteq
    \mathcal{W}^{\mathrm{R}}(S).
\]
\end{corollary}

The inclusion can be strict, as illustrated in \Cref{fig:staking-restaking-example}(a). More generally, over-coverage strictly reduces the restaking load contributed by a positive commitment:
if $w_{ij}>0$ and $W_j>r_j$, then
\[
    r_j\frac{w_{ij}}{W_j}<w_{ij}.
\]

\subsection{Feasibility Without Incentive Constraints}
\label{subsec:market-set-feasibility-equivalence}

We next ignore incentive and equilibrium constraints and ask only whether there exists a commitment profile that safely supports a given market set.
Let
\[
    \Sigma:=\sum_{i\in V}\sigma_i.
\]

\begin{proposition}[Batch market-set feasibility]
\label[proposition]{prop:market-set-feasibility-equivalence}
For every $S\subseteq\mathcal{B}_t$, the following are equivalent: (i) there exists a staking-safe allocation $(S,w)$; (ii) there exists a restaking-safe allocation $(S,w)$; (iii) $\sum_{j\in S}r_j\le\Sigma$.
\end{proposition}

\begin{proof}
If $(S,w)$ is staking-safe, then
\[
    \sum_{j\in S}r_j
    \le\sum_{j\in S}W_j
    =\sum_{i\in V}\Lambda_i^{\mathrm{S}}(S,w)
    \le\Sigma.
\]
Hence (i) implies (iii).

If $(S,w)$ is restaking-safe, then
\[
    \sum_{j\in S}r_j
    =\sum_{i\in V}\Lambda_i^{\mathrm{R}}(S,w)
    \le\Sigma,
\]
so (ii) implies (iii).

Conversely, suppose
\[
    \sum_{j\in S}r_j\le\Sigma
\]
and define
\[
    \widetilde w_{ij}
    :=\frac{\sigma_i}{\Sigma}r_j,
    \qquad i\in V,\ j\in S.
\]
Since $r_j\le\sum_{k\in S}r_k\le\Sigma$,
\[
    0\le \widetilde w_{ij}\le\sigma_i.
\]
Moreover,
\[
    \widetilde W_j
    =\sum_{i\in V}\widetilde w_{ij}
    =r_j,
\]
and
\[
    \sum_{j\in S}\widetilde w_{ij}
    =\frac{\sigma_i}{\Sigma}\sum_{j\in S}r_j
    \le\sigma_i.
\]
Thus $(S,\widetilde w)$ is staking-safe.

Since $\widetilde W_j=r_j$ for every $j\in S$,
\[
    \Lambda_i^{\mathrm{R}}(S,\widetilde w)
    =
    \Lambda_i^{\mathrm{S}}(S,\widetilde w)
    \le\sigma_i,
\]
so $(S,\widetilde w)$ is also restaking-safe.
\end{proof}

\begin{corollary}[Feasible market sets]
\label[corollary]{cor:feasible-market-set-equivalence}
Without incentive or equilibrium constraints, staking and restaking support exactly the same market subsets:
\[
    \left\{
        S\subseteq\mathcal{B}_t:
        \sum_{j\in S}r_j\le\sum_{i\in V}\sigma_i
    \right\}.
\]
\end{corollary}

\begin{remark}
Absent incentive constraints, batch-level safety feasibility reduces to the aggregate condition
\[
    \sum_{j\in S}r_j\le\sum_{i\in V}\sigma_i.
\]
Hence, subsequent differences between staking and restaking arise from restrictions on implementable commitment profiles induced by strategic participation and reward mechanisms.
\end{remark}

\section{Single Market Reward Comparison}
\label[appendix]{app:single-market-reward-mapping}

Consider a batch containing a single market $j$, i.e., $S=\{j\}$. Each voter $i$ may commit any amount $w_{ij}\in[0,\sigma_i]$.
For readability, write $c_i:=c_{ij}$. For a single market, the admission certificate reduces to the aggregate requirement $W_j\geq r_j$.

Indeed, once this condition holds, each voter's proportional load satisfies
\[
    r_j\frac{w_{ij}}{W_j}
    \leq
    w_{ij}
    \leq
    \sigma_i.
\]
Thus, admission is monotone in the market's total committed stake.

\begin{proposition}[Single-market equilibrium mapping]
\label[proposition]{prop:single-market-reward-mapping}
Fix a single market $j$.

\begin{enumerate}[label=(\roman*)]
    \item Suppose a per-stake reward rate $b_j$ supports an equilibrium
    $w^{A}$ with total committed stake
    \[
        W^{A}:=\sum_{i\in V}w_{ij}^{A}>0.
    \]
    Then the fixed reward pool
    \[
        R_j:=b_jW^{A}
    \]
    supports the same equilibrium commitment profile, with the same
    total committed stake and the same total reward expenditure.

    \item Suppose a fixed reward pool $R_j$ supports an equilibrium
    $w^{P}$ with total committed stake
    \[
        W^{P}:=\sum_{i\in V}w_{ij}^{P}>0.
    \]
    Then the per-stake reward rate
    \[
        b_j:=\frac{R_j}{W^{P}}
    \]
    supports an equilibrium $w^{E}$ whose total committed stake
    $W^{E}$ satisfies
    \[
        W^{E}\geq W^{P}.
    \]
    Hence, if $w^{P}$ passes the single-market admission certificate,
    then so does $w^{E}$. The per-stake rule pays
    \[
        b_jW^{E}\geq R_j,
    \]
    and this inequality may be strict.
\end{enumerate}
\end{proposition}

The first direction sets the fixed pool equal to the total payment made by the per-stake rule at the original equilibrium. When $b_j>0$, every participating voter commits its full available stake under the per-stake rule, and receives exactly the same reward under the fixed pool. When $b_j=0$, every participating voter must have zero participation cost, so the same profile is also an equilibrium under the zero fixed pool. A nonparticipating voter cannot profitably enter in either case.

For the reverse direction, we set the per-stake rate equal to the average fixed-pool reward rate at the original equilibrium. Every original participant then remains willing to participate at its full available stake. Additional voters may also enter because their participation no longer dilutes the rewards of others. Aggregate commitment therefore weakly increases, but the protocol may pay more in total.

\begin{proof}
We prove the two directions separately.

\parhead{From per-stake rewards to a fixed pool}
Consider a per-stake equilibrium $w^{A}$ with total commitment
\[
    W^{A}=\sum_{i\in V}w_{ij}^{A}>0.
\]
Set
\[
    R_j:=b_jW^{A}.
\]

Suppose first that $b_j>0$. Conditional on participation, voter $i$'s payoff
\[
    b_jw_{ij}-c_i
\]
is strictly increasing in $w_{ij}$. Hence every participating voter commits its full available stake:
\[
    w_{ij}^{A}=\sigma_i
    \qquad\text{whenever }w_{ij}^{A}>0.
\]
The equilibrium conditions therefore imply
\[
    b_j\sigma_i\geq c_i,
    \qquad w_{ij}^{A}>0,
\]
and
\[
    b_j\sigma_k\leq c_k,
    \qquad w_{kj}^{A}=0.
\]

At the original commitment profile, each participating voter $i$ receives
\[
    R_j\frac{\sigma_i}{W^{A}}
    =
    b_jW^{A}\frac{\sigma_i}{W^{A}}
    =
    b_j\sigma_i.
\]
Thus, its fixed-pool payoff is identical to its per-stake payoff and is nonnegative.

Fix the commitments of all other voters. For a participating voter $i$, the fixed-pool reward from a positive commitment $x\in(0,\sigma_i]$ is
\[
    R_j
    \frac{x}{W^{A}-\sigma_i+x}.
\]
This expression is nondecreasing in $x$, so $\sigma_i$ is a best response among positive commitments. Since its payoff at $\sigma_i$ is nonnegative, abstaining is not a profitable deviation.

Now consider a nonparticipant $k$. If it enters with any $x\in(0,\sigma_k]$, its reward is
\[
    R_j\frac{x}{W^{A}+x}
    =
    b_jW^{A}\frac{x}{W^{A}+x}
    <
    b_jx
    \leq
    b_j\sigma_k
    \leq
    c_k.
\]
Thus, no nonparticipant can profitably enter.

It remains to consider $b_j=0$. Then $R_j=0$. Every voter participating in the per-stake equilibrium must have $c_i=0$, and hence obtains zero payoff from any positive commitment under both reward rules. Any nonparticipant has $c_k\geq0$ and therefore cannot profit from entry. Thus, the same commitment profile is also an equilibrium under the fixed-pool rule.

Finally, the fixed-pool payment is
\[
    R_j=b_jW^{A},
\]
which is exactly the total payment under the per-stake rule.

\parhead{From a fixed pool to per-stake rewards}
Now consider a fixed-pool equilibrium $w^{P}$ with total commitment
\[
    W^{P}
    =
    \sum_{i\in V}w_{ij}^{P}
    >
    0.
\]
Recall that the per-stake reward rate is $b_j=R_j/W^{P}$. Every voter $i$ with $w_{ij}^{P}>0$ must weakly prefer participation to abstention. Therefore,
\[
    R_j\frac{w_{ij}^{P}}{W^{P}}
    \geq
    c_i
    \iff
    b_j w_{ij}^{P}\geq c_i.
\]
Since $\sigma_i \geq w_{ij}^{P}$, we obtain $b_j\sigma_i\geq c_i$ for every voter participating in the fixed-pool equilibrium.

Define
\[
    A^{E}
    :=
    \left\{
        i\in V:
        b_j\sigma_i\geq c_i
    \right\}.
\]
Consider the per-stake commitment profile
\[
    w_{ij}^{E}
    =
    \begin{cases}
        \sigma_i, & i\in A^{E},\\
        0, & i\notin A^{E}.
    \end{cases}
\]
For every $i\in A^{E}$, committing $\sigma_i$ yields a nonnegative payoff and maximizes its payoff among positive commitments. For every $k\notin A^{E}$, any positive commitment $x\leq\sigma_k$ gives
\[
    b_jx-c_k
    \leq
    b_j\sigma_k-c_k
    <
    0.
\]
Thus, $w^{E}$ is a per-stake equilibrium.

Every voter participating in the original fixed-pool equilibrium belongs to $A^{E}$. Consequently,
\[
    W^{E}
    :=
    \sum_{i\in A^{E}}\sigma_i
    \geq
    \sum_{i:w_{ij}^{P}>0}\sigma_i
    \geq
    \sum_{i\in V}w_{ij}^{P}
    =
    W^{P}.
\]
Because the single-market admission certificate is monotone in aggregate commitment, admission of $w^{P}$ implies admission of $w^{E}$.
Finally, the total per-stake payment is
\[
    b_jW^{E}
    \geq
    b_jW^{P}
    =
    R_j.
\]
The inequality is strict whenever the per-stake rule attracts additional commitment beyond $W^{P}$.
\end{proof}

\begin{corollary}[Single-market equilibrium-expenditure dominance]
\label{cor:single-market-reward-dominance}
For an admission-feasible single market, the minimum fixed-pool reward expenditure required to support an equilibrium whose commitment profile passes the admission certificate is no greater than the corresponding minimum per-stake reward expenditure.
\end{corollary}

\begin{proof}
Any per-stake equilibrium whose commitment profile passes the admission certificate can be mapped, by \Cref{prop:single-market-reward-mapping}(i), to a fixed-pool equilibrium with the same profile and total payment. Taking the minimum over all per-stake rewards that support such an equilibrium gives the result.
\end{proof}

\parhead{Example: Strict reward savings under a fixed pool}
Consider a single-market batch $S=\{j\}$ with admission stake requirement $r_j=10$.
There are two voters, $A$ and $B$, whose available stakes and verification costs are
\[
    \sigma_A=100,
    \qquad
    c_{Aj}=0.9,
\]
and
\[
    \sigma_B=1000,
    \qquad
    c_{Bj}=5.
\]
Each voter may commit any amount in its interval $[0,\sigma_i]$. We use the tie-breaking convention that a voter participates when indifferent.
Under a per-stake reward rate $b_j$, voter $i$ participates at its full available stake if $b_j\cdot\sigma_i\geq c_{ij}$.
The two break-even reward rates are
\[
    \frac{c_{Aj}}{\sigma_A}=0.009,
    \qquad
    \frac{c_{Bj}}{\sigma_B}=0.005.
\]
Hence, the lowest per-stake rate that induces any positive participation is
\(
    b_j=0.005.
\)
At this rate, only voter $B$ participates, yielding
\(
    W_j=1000\geq r_j.
\)
The market is therefore admitted, but the protocol pays
\[
    C_j^{\mathrm{ps}}
    =
    b_jW_j
    =
    0.005\cdot 1000
    =
    5.
\]

Now consider a fixed reward pool $R_j=1$ and the commitment profile
\[
    w_{Aj}=100,
    \qquad
    w_{Bj}=0.
\]
Voter $A$ receives the entire pool and obtains payoff
\[
    1-0.9=0.1>0,
\]
so participation is a best response.
If voter $B$ enters with commitment $x\in(0,1000]$, its payoff is
\[
    \frac{x}{100+x}-5.
\]
This expression is increasing in $x$, and even at $x=1000$ it is
\[
    \frac{1000}{1100}-5
    <
    0.
\]
Thus, voter $B$ does not benefit from entering. The profile is therefore an equilibrium. Its total commitment is
\[
    W_j=100\geq r_j,
\]
while the protocol pays only
\[
    C_j^{\mathrm{pool}}=R_j=1.
\]

Consequently, if $C_j^{\mathrm{ps},\star}$ and
$C_j^{\mathrm{pool},\star}$ denote the minimum expenditures required to support a market-admitting equilibrium under the two reward rules, then
\[
    C_j^{\mathrm{pool},\star}
    \leq
    1
    <
    5
    =
    C_j^{\mathrm{ps},\star}.
\]
Thus, the fixed-pool rule can require strictly less reward expenditure than the per-stake rule.

This example concerns the reward expenditure required to support a market-admitting equilibrium. It does not imply that every equilibrium induced by the fixed-pool rule passes the admission certificate.

\section{Omitted Proofs}
\label[appendix]{sec:proof}

\subsection{Economic-Security Guarantee Proof}
\label[appendix]{app:proof-restaking-economic-security}

\parhead{Restatement of the guarantee}
Recall that the admission stake requirement of market $j$ is $r_j=\frac{\pi_j}{\alpha_j}$.
A pair $(S,w)$ passes the admission certificate if
\[
    W_j\geq r_j,
    \qquad \forall j\in S,
\]
and
\[
    \Lambda_i^R(S,w)
    =
    \sum_{j\in S}r_j\frac{w_{ij}}{W_j}
    \leq \sigma_i,
    \qquad \forall i\in V.
\]
The first condition ensures that $r_j/W_j\leq 1$ for every admitted market, while the second bounds the total restaking load assigned to each voter by its slashable stake.

\Cref{thm:restaking-economic-security} states that these two conditions are sufficient for economic security. In particular, for every attacking-stake profile $a$ satisfying $0\leq a_{ij}\leq w_{ij}$, we show that the total manipulation gain of the successfully corrupted markets is no greater than the coalition's attack cost:
\[
    \sum_{j\in\mathcal{C}_S(a;w)}\pi_j
    \leq
    \sum_{i\in V}c_i^{\mathrm{att}}(a;w).
\]
It then follows directly from \Cref{def:coalitional-attack} that $U_S^{\mathrm{att}}(a;w)\leq 0$.

\begin{proof}[Proof of \Cref{thm:restaking-economic-security}]
Fix any attacking-stake profile $a$ satisfying $0\leq a_{ij}\leq w_{ij}$ for all $i\in V$ and $j\in S$, and let
\[
    C := \mathcal{C}_S(a;w)
\]
denote the set of successfully corrupted markets. If $C=\emptyset$, then $U_S^{\mathrm{att}}(a;w)=0$, so the claim holds immediately. Suppose therefore that $C\neq\emptyset$.

For every corrupted market $j\in C$, by definition of corruption,
\[
    \sum_{i\in V}a_{ij}
    \geq
    \alpha_j W_j.
\]
Since $r_j=\pi_j/\alpha_j$, we have
\[
    \pi_j =
    \alpha_j r_j
    \leq
    \frac{r_j}{W_j}
    \sum_{i\in V}a_{ij}
    =
    \sum_{i\in V}
    \frac{r_j}{W_j}a_{ij}.
\]
Summing over all corrupted markets gives
\[
    \sum_{j\in C}\pi_j
    \leq
    \sum_{j\in C}
    \sum_{i\in V}
    \frac{r_j}{W_j}a_{ij}
    =
    \sum_{i\in V}
    \sum_{j\in C}
    \frac{r_j}{W_j}a_{ij}.
\]

For each voter $i$, define its assigned attack charge as
\[
    q_i
    :=
    \sum_{j\in C}
    \frac{r_j}{W_j}a_{ij}.
\]
We show that this charge is no greater than voter $i$'s attack cost:
\[
    q_i
    \leq
    c_i^{\mathrm{att}}(a;w).
\]

First, because $a_{ij}\leq w_{ij}$ and condition (ii) of the admission
certificate holds,
\[
    q_i =
    \sum_{j\in C}
    \frac{r_j}{W_j}a_{ij}
    \leq
    \sum_{j\in S}
    \frac{r_j}{W_j}w_{ij}
    =
    \Lambda_i^R(S,w)
    \leq
    \sigma_i.
\]

Second, condition (i) of the admission certificate gives $W_j\geq r_j$ for every $j\in S$, and hence $r_j/W_j\leq 1$. Therefore,
\[
q_i = \sum_{j\in C} \frac{r_j}{W_j}a_{ij} \leq \sum_{j\in C}a_{ij}.
\]
Combining these two bounds yields
\[
    q_i
    \leq
    \min\left\{
        \sigma_i,
        \sum_{j\in C}a_{ij}
    \right\}
    =
    c_i^{\mathrm{att}}(a;w).
\]
Finally, summing the assigned charges across voters gives
\[
    \sum_{j\in C}\pi_j
    \leq
    \sum_{i\in V}q_i
    \leq
    \sum_{i\in V}c_i^{\mathrm{att}}(a;w).
\]
Therefore,
\[
    U_S^{\mathrm{att}}(a;w)
    =
    \sum_{j\in C}\pi_j
    -
    \sum_{i\in V}c_i^{\mathrm{att}}(a;w)
    \leq
    0.
\]
Since $a$ was arbitrary, no attacking-stake profile is profitable.
Therefore, $(S,w)$ is economically secure.
\end{proof}

\subsection{Proofs for Reward-Setting Obstacles}
\label[appendix]{app:multi-market-hardness-proofs}

This section proves the three claims underlying the reward-setting difficulties in \Cref{subsec:new-multimarket-obstacles}. We first show that deciding whether fixed rewards support a pure-strategy equilibrium that passes the admission certificate is NP-hard even for a single market. We then show that, with multiple markets, the set of one voter's commitments that make the resulting joint profile pass the admission certificate can be nonconvex. Finally, we show that equilibrium decomposes market by market under fixed-pool rewards, whereas admission need not.

\begin{proof}[Proof of \Cref{prop:reward-setting-combinatorial-hardness}]
We reduce from \textsc{Subset Sum}.
Let the instance consist of positive integers $q_1,\ldots,q_n$ and target $B$; items with $q_i>B$ may be discarded. 

Construct a single market with admission requirement $r=B$ and fixed reward pool $R=B$. Voter $i$ has stake $\sigma_i=q_i$ and fixed participation cost $c_i=q_i$.

Let $W$ be the total commitment. If voter $i$ participates with $w_i>0$, its payoff is
\[
    B\frac{w_i}{W}-q_i.
\]
For fixed $W_{-i}>0$, this expression is strictly increasing in $w_i$, so every positive best response uses the full commitment $w_i=q_i$. If $W_{-i}=0$, an admitted profile is possible only when $q_i=B$, which already gives an exact subset-sum solution.

Suppose first that some subset $A$ satisfies
\[
\sum_{i\in A}q_i=B.
\]
Let each $i\in A$ commit $q_i$ and all other voters commit zero. Every participant obtains zero payoff:
\[
B\frac{q_i}{B}-q_i=0.
\]

If $A$ contains another participant, any smaller positive commitment gives strictly lower payoff. If $A=\{i\}$, then $q_i=B$ and every positive commitment gives the same zero payoff, so $q_i$ is still a best response. An outsider $k$ obtains at most
\[
    B\frac{q_k}{B+q_k}-q_k<0
\]
by entering. Hence the profile is a pure-strategy equilibrium and passes the admission certificate.

Conversely, suppose an admitted pure-strategy equilibrium exists, and let $A$ be its set of positive participants.  If $|A|=1$, admission requires the sole participant to have $q_i=B$, giving a subset-sum solution. Otherwise, every participant commits $q_i$, so
\[
    W=\sum_{i\in A}q_i.
\]

Admission requires $W\ge B$. For every participant, remaining in the equilibrium must be at least as good as exiting, so
\[
    B\frac{q_i}{W}-q_i\ge0,
\]
which implies $W\le B$. Thus $W=B$, and $A$ solves the original \textsc{Subset Sum} instance. The single-market equilibrium-support problem is therefore NP-hard.

The construction also satisfies the restaking capacity condition: when $W=B$, voter $i$'s restaking load is
\[
    B\frac{q_i}{W}=q_i=\sigma_i.
\]

\end{proof}

\begin{proof}[Proof of \Cref{prop:restaking-nonconvex-admission}]
It suffices to exhibit fixed commitments of the other voters and two commitment vectors of one voter such that both resulting joint profiles pass the admission certificate, while their midpoint does not.

Consider two voters \(i,k\) and two markets \(S=\{1,2\}\). Let
\[
    \sigma_i=1,
    \qquad
    \sigma_k=10,
    \qquad
    r_1=r_2=\frac{51}{10},
\]
and fix voter \(k\)'s commitments as
\(
    w_k=(5,5).
\)

Consider the two commitment vectors of voter $i$
\[
    x=\left(1,\frac{1}{10}\right),
    \qquad
    y=\left(\frac{1}{10},1\right).
\]

At $x$, both markets satisfy the coverage requirement:
\[
    W_1=6\ge\frac{51}{10},
    \qquad
    W_2=\frac{51}{10}.
\]

Voter $i$'s cumulative restaking load is
\[
    \frac{51}{10}\times\frac{1}{6}
    +
    \frac{51}{10}\times\frac{1/10}{5+1/10}
    =
    \frac{19}{20}
    <1,
\]
while voter $k$'s load is
\[
    \frac{51}{10}\times\frac{5}{6}
    +
    \frac{51}{10}\times\frac{5}{5+1/10}
    =
    \frac{37}{4}
    <10.
\]

Hence the resulting joint profile passes the admission certificate. By symmetry, the same holds at $y$.

Now consider their midpoint
\[
m=\frac{x+y}{2}=\left(\frac{11}{20},\frac{11}{20}\right).
\]

Both markets remain covered, since
\[
W_1=W_2=\frac{111}{20}>\frac{51}{10}.
\]

Voter $k$'s cumulative load remains below its capacity:
\[
    2\times
    \frac{51}{10}\times
    \frac{5}{5+11/20}
    =
    \frac{340}{37}
    <10.
\]

However, voter $i$'s cumulative load is
\[
    2\times
    \frac{51}{10}\times
    \frac{11/20}{5+11/20}
    =
    \frac{187}{185}
    >1.
\]

Thus the joint profile at $m$ fails the admission certificate, although the profiles at $x$ and $y$ both pass it. Therefore, even after fixing all other voters' commitments, the set of commitments of one voter for which the resulting joint profile passes the admission certificate need not be convex.

\end{proof}

\begin{proof}[Proof of \Cref{prop:restaking-fixed-pool-generalized-aggregative}]
Under fixed market-level reward pools, voter $i$'s action set is the product
\[
    [0,\sigma_i]^S
    =
    \prod_{j\in S}[0,\sigma_i],
\]
and its participation payoff is additive across markets. Moreover, the payoff from market $j$ depends only on $w_{ij}$ and the aggregate commitment $W_{-i,j}$ of the other voters in that market.
Suppose first that a commitment profile is a pure-strategy equilibrium in every market separately. Then, for every voter $i$, changing $w_{ij}$ in any single market cannot increase its payoff from that market. Since the total payoff is the sum of these market-specific payoffs, no deviation across several markets can increase voter $i$'s total payoff. Hence the profile is a pure-strategy equilibrium of the multi-market participation game.

Conversely, if the profile is not a pure-strategy equilibrium in some market $j$, then some voter has a profitable deviation in that market. Changing only its commitment to market $j$ while keeping all other market commitments unchanged gives the same payoff improvement in the multi-market game. Hence the original profile cannot be a pure-strategy equilibrium of the multi-market participation game. Therefore, equilibrium in the multi-market participation game is equivalent to equilibrium in every market separately.

It remains to show that admission does not share this decomposition. We give a counterexample with one voter $i$ and two markets $S=\{1,2\}$, where
\[
\sigma_i=1, \qquad r_1=r_2=\frac{3}{5}.
\]

Let both markets offer a positive fixed reward and let the participation costs be zero. For either market considered alone, $w_{ij}=1$ is a pure-strategy equilibrium and the resulting single-market profile passes the admission certificate: the market is covered, and the voter's restaking load is
\[
    \frac{3}{5}\le 1.
\]

Combining the two commitments gives $w_i=(1,1)$.
By the equilibrium decomposition above, this profile is a pure-strategy equilibrium of the two-market participation game. Both markets are covered, but the voter's cumulative restaking load is
\[
    \frac{3}{5}+\frac{3}{5}
    =
    \frac{6}{5}
    >1.
\]

Hence the combined profile fails the admission certificate for $S=\{1,2\}$. Thus, equilibrium commitments that support admission in each market separately need not support admission when combined across markets.

\end{proof}

\subsection{Proofs of Greedy Equilibrium Construction}
\label[appendix]{app:greedy-equilibrium-proof}

\parhead{Restatement of \Cref{thm:fixed-pool-greedy-ne}} For a single-market settlement batch with fixed-pool rewards, the profile $w^G$ returned by \Cref{alg:single-market-greedy} is a pure-strategy equilibrium. If $W^G\ge r_j$, then $(\{j\},w^G)$ also passes the admission certificate.

\begin{proof}[Proof of \Cref{thm:fixed-pool-greedy-ne}]
We first show that no selected voter has a profitable deviation.
If $A^G=\emptyset$, this part is vacuous. Otherwise, let $(m)$ be the last voter added by the scan. Since no voter is selected after $(m)$, its successful entry test is evaluated at the final aggregate commitment $W^G$, giving
\[
    \frac{R_j}{W^G}
    \ge
    \eta_{(m)j}.
\]
Because voters are scanned in nondecreasing order of $\eta_{ij}$, every selected voter $i\in A^G$ satisfies $\eta_{ij}\le\eta_{(m)j}$. Hence
\[
    \frac{R_j}{W^G}
    \ge
    \eta_{ij}
    =
    \frac{c_{ij}}{\sigma_i},
\]
and therefore
\[
    R_j\frac{\sigma_i}{W^G}
    \ge
    c_{ij}.
\]
Thus, every selected voter obtains a nonnegative payoff from participation. Moreover, for fixed commitments of the other voters, the gross reward $R_jx/(W_{-i,j}+x)$ is nondecreasing in $x$. Hence $\sigma_i$ is a best response among positive commitments, so no selected voter benefits from reducing its commitment or exiting.

We next show that no skipped voter has a profitable deviation. Let $k\notin A^G$, and let $W_k$ denote the aggregate commitment when voter $k$ was scanned. Since $k$ failed its entry test,
\[
    \frac{R_j}
         {W_k+\sigma_k}
    <
    \eta_{kj}.
\]
The aggregate commitment can only increase after voter $k$ is scanned, so $W_k\le W^G$ and
\[
    \frac{R_j}
         {W^G+\sigma_k}
    \le
    \frac{R_j}
         {W_k+\sigma_k}
    <
    \eta_{kj}.
\]
It follows that
\[
    R_j
    \frac{\sigma_k}
         {W^G+\sigma_k}
    <
    c_{kj}.
\]
For fixed $W^G$, the gross reward from entry is nondecreasing in voter $k$'s commitment. Hence, if entry at $\sigma_k$ is unprofitable, every smaller positive commitment is also unprofitable. Thus, voter $k$ prefers not to enter.

No selected voter wants to reduce its commitment or exit, and no skipped voter wants to enter. Therefore, $w^G$ is a pure-strategy equilibrium.

Suppose now that $W^G\ge r_j$. Condition~(i) of the admission certificate (\Cref{def:safe-allocation}) holds immediately. For every selected voter $i$,
\[
    \Lambda_i^R(\{j\},w^G)
    =
    r_j\frac{\sigma_i}{W^G}
    \le
    \sigma_i,
\]
while every unselected voter has zero cumulative restaking load.
Hence Condition~(ii) also holds, and $(\{j\},w^G)$ passes the admission certificate.

Taking $H=V$ and $w_H^\star=w^G$, the admission condition in \Cref{def:settlement-liveness} follows from the argument above.
Moreover, since $H=V$, there are no voters outside $H$, and the robust best-response condition reduces to the standard equilibrium condition. Because $w^G$ is a pure-strategy equilibrium, this condition also holds. Therefore, the reward rule induced by $R_j$ provides settlement liveness for market $j$.
\end{proof}

\parhead{Existence of a sufficient reward pool} For every admission-feasible single market $j$, a finite reward pool sufficient for the greedy construction exists.

\begin{proof}
For a single market, admission feasibility implies
\[
    r_j \le \sum_{i\in V}\sigma_i.
\]
Consider any reward pool satisfying
\[
    R_j
    \ge
    \left(\sum_{i\in V}\sigma_i\right)
    \max_{i:\sigma_i>0}\frac{c_{ij}}{\sigma_i}.
\]
We show that every voter with positive available stake passes the greedy entry condition. When voter $i$ is scanned, let $W$ denote the aggregate commitment of the voters selected so far. Then
\[
    W+\sigma_i
    \le
    \sum_{k\in V}\sigma_k,
\]
and hence
\[
    \frac{R_j}{W+\sigma_i}
    \ge
    \frac{R_j}{\sum_{k\in V}\sigma_k}
    \ge
    \frac{c_{ij}}{\sigma_i}.
\]
Thus voter $i$ passes the greedy entry condition. Therefore, every voter with $\sigma_i>0$ is selected, and the final aggregate commitment satisfies
\[
    W^G=\sum_{i\in V}\sigma_i\ge r_j.
\]
Hence the stated finite reward pool is sufficient.
\end{proof}

\subsection{Monotonicity of the Greedy Construction}
\label[appendix]{app:greedy-monotonicity}

We establish the monotonicity property used by the reward search in \Cref{sec:single-market-greedy-construction}. Fix a market $j$. As in \Cref{sec:single-market-greedy-construction}, voters with $\sigma_i=0$ are omitted. Order the remaining voters as in \Cref{alg:single-market-greedy}, using the fixed tie-breaking rule, so that
\[
    \eta_{(1)j}\le \eta_{(2)j}\le\cdots\le\eta_{(n)j},
    \qquad
    \eta_{ij}:=\frac{c_{ij}}{\sigma_i}.
\]
For a reward pool $R$, let $W_\ell(R)$ denote the aggregate commitment after the first $\ell$ voters have been processed by the greedy scan. Thus,
\[
    W_0(R)=0,
    \qquad
    W_n(R)=W^G(R).
\]

\begin{lemma}[Monotonicity of the greedy commitment]
\label[lemma]{lem:greedy-monotonicity}
Let $W^G(R)$ denote the aggregate commitment returned by
\Cref{alg:single-market-greedy} under reward pool $R$. For any
$0\le R\le R'$,
\[
    W^G(R)\le W^G(R').
\]
\end{lemma}

\begin{proof}
Apply the greedy construction under reward pools $R$ and $R'$ using the same fixed voter ordering. Let
\[
    q:=
    \max\Bigl(
        \bigl\{
            \ell\in\{1,\ldots,n\}:
            \eta_{(\ell)j}=0
        \bigr\}
        \cup\{0\}
    \Bigr).
\]
Because the voters are ordered by nondecreasing break-even rate, the first $q$ voters are exactly those with zero break-even rate. Each of these voters is selected under both reward pools. Hence,
\[
    W_q(R)
    =
    W_q(R')
    =
    \sum_{h=1}^{q}\sigma_{(h)}.
\]
If $q=n$, the two scans return the same aggregate commitment and the claim follows immediately. For every $\ell>q$, define the acceptance threshold under the smaller reward pool by
\[
    T_\ell:=\frac{R}{\eta_{(\ell)j}}.
\]
Since the positive break-even rates are nondecreasing,
\[
    T_{q+1}\ge T_{q+2}\ge\cdots\ge T_n.
\]
Under reward pool $R$, voter $(\ell)$ is selected exactly when
\[
    W_{\ell-1}(R)+\sigma_{(\ell)}\le T_\ell.
\]

We prove by induction over $\ell=q,\ldots,n$ that
\[
    W_\ell(R')\ge W_\ell(R).
\]
We additionally prove that, for every $\ell>q$, strict inequality
implies
\[
    W_\ell(R')>T_\ell.
\]
The base case $\ell=q$ follows from
$W_q(R')=W_q(R)$.

Fix $\ell>q$ and suppose the claims hold after the first
$\ell-1$ voters have been processed.

First, suppose both scans make the same decision. They either both
add $\sigma_{(\ell)}$ or both add nothing, so
\[
    W_\ell(R')\ge W_\ell(R).
\]
If the inequality after this step is strict, then it was already
strict before this step. Because the two aggregates are equal at
step $q$, this implies $\ell-1>q$. By the induction hypothesis,
\[
    W_{\ell-1}(R')>T_{\ell-1}\ge T_\ell.
\]
Since $W_\ell(R')\ge W_{\ell-1}(R')$, it follows that
\[
    W_\ell(R')>T_\ell.
\]

Next, suppose the scan under $R$ rejects voter $(\ell)$ while the scan under $R'$ accepts it. Rejection under $R$ implies
\[
    W_{\ell-1}(R)+\sigma_{(\ell)}>T_\ell.
\]
Therefore,
\[
\begin{aligned}
    W_\ell(R')
    &=
    W_{\ell-1}(R')+\sigma_{(\ell)}\\
    &\ge
    W_{\ell-1}(R)+\sigma_{(\ell)}
    >
    T_\ell.
\end{aligned}
\]
Moreover, because the scan under $R$ rejects the voter,
\[
\begin{aligned}
    W_\ell(R')
    &\ge
    W_{\ell-1}(R)+\sigma_{(\ell)}\\
    &>
    W_{\ell-1}(R)
    =
    W_\ell(R).
\end{aligned}
\]
Thus both induction claims hold.

Finally, suppose the scan under $R$ accepts voter $(\ell)$ while the scan under $R'$ rejects it. The two scans cannot have the same aggregate commitment before processing this voter. Indeed, if
\[
    W_{\ell-1}(R')=W_{\ell-1}(R),
\]
then acceptance under $R$ would give
\[
    W_{\ell-1}(R')+\sigma_{(\ell)}
    \le
    \frac{R}{\eta_{(\ell)j}}
    \le
    \frac{R'}{\eta_{(\ell)j}},
\]
so the scan under $R'$ would also accept voter $(\ell)$.

Hence,
\[
    W_{\ell-1}(R')>W_{\ell-1}(R).
\]
Since the aggregates are equal at step $q$, this also implies $\ell-1>q$. By the induction hypothesis,
\[
    W_{\ell-1}(R')>T_{\ell-1}\ge T_\ell.
\]
Meanwhile, acceptance under $R$ implies
\[
    W_\ell(R)
    =
    W_{\ell-1}(R)+\sigma_{(\ell)}
    \le
    T_\ell.
\]
Because the scan under $R'$ rejects the voter,
\[
    W_\ell(R')
    =
    W_{\ell-1}(R')
    >
    T_\ell
    \ge
    W_\ell(R).
\]
Thus both induction claims again hold. Therefore,
\[
    W_\ell(R')\ge W_\ell(R)
\]
after every step of the scan. Taking $\ell=n$ gives
\[
    W^G(R')\ge W^G(R).
\]
\end{proof}

\subsection{Proof of the Protected-Layer Settlement Liveness Guarantee}
\label[appendix]{app:proof-robust-protected-voter-liveness}

\parhead{Restatement of \Cref{thm:robust-protected-voter-liveness}}
Recall that the protected-layer construction selects a protected voter set $H_\theta$ satisfying
\[
    B_H(\theta)\ge \mathcal{L}(S),
\]
and assigns every protected voter \(i\in H_\theta\) the reference commitment
\[
    w_{ij}^\star=\sigma_i,
    \qquad j\in S.
\]
The theorem states that, for every commitment profile
\[
    w_{-H}\in\prod_{k\notin H}[0,\sigma_k]^S,
\]
the joint profile $(w_H^\star,w_{-H})$ passes the admission certificate. Moreover, each protected voter's reference commitment is a best response whenever the other protected voters follow the
reference profile. Hence, the constructed reward pools provide settlement liveness for $S$.

\begin{proof}[Proof of
\Cref{thm:robust-protected-voter-liveness}]
Since $S$ is admission-feasible, there exists a commitment profile $\widetilde w$ that passes the admission certificate. Therefore,
\[
    L(S)
    =
    \sum_{j\in S}r_j
    =
    \sum_{i\in V}\Lambda_i^R(S,\widetilde w)
    \le
    \sum_{i\in V}\sigma_i.
\]
Hence, once the threshold is lowered to include every voter with positive available stake,
\[
    B_H(\theta)=\sum_{i\in V}\sigma_i\ge L(S),
\]
so the construction finds a qualifying threshold for every admission-feasible target batch.

Fix the selected threshold $\theta$, and write
\[
    H:=H_\theta,\qquad
    B_H:=B_H(\theta),\qquad
    B_L:=B_L(\theta).
\]
For every $i\in H$ and $j\in S$, the reference profile sets
\[
    w_{ij}^\star=\sigma_i.
\]
Fix any commitment profile
\[
    w_{-H}\in\prod_{k\notin H}[0,\sigma_k]^S,
\]
and let
\[
    e_j:=\sum_{k\notin H}w_{kj}.
\]
For every market $j\in S$,
\[
    W_j=B_H+e_j\ge B_H\ge L(S)\ge r_j,
\]
so Condition~(i) in \Cref{def:safe-allocation} holds.

For every protected voter $i\in H$,
\[
    \Lambda_i^R(S,w)
    =
    \sum_{j\in S}
    r_j\frac{\sigma_i}{B_H+e_j}
    \le
    \frac{\sigma_i}{B_H}\sum_{j\in S}r_j
    =
    \sigma_i\frac{L(S)}{B_H}
    \le
    \sigma_i.
\]
For every non-protected voter $k\notin H$,
\[
    \Lambda_k^R(S,w)
    =
    \sum_{j\in S}
    r_j\frac{w_{kj}}{B_H+e_j}
    \le
    \frac{\sigma_k}{B_H}\sum_{j\in S}r_j
    =
    \sigma_k\frac{L(S)}{B_H}
    \le
    \sigma_k.
\]
Hence Condition~(ii) also holds, and $(S,(w_H^\star,w_{-H}))$ passes the admission certificate.

We next establish the best-response condition.
Fix a protected voter $i\in H$ and an arbitrary non-protected profile $w_{-H}$, and hold all other protected voters at their reference commitments. As above, let
\[
    e_j:=\sum_{k\notin H}w_{kj},
\]
and define
\[
    d_{ij}:=B_H-\sigma_i+e_j.
\]
If voter $i$ commits $x\in[0,\sigma_i]$ to market $j$, its gross reward is
\[
    g_{ij}(x)
    :=
    R_j^{\mathrm{robust}}(\theta)
    \frac{x}{x+d_{ij}},
\]
with $g_{ij}(0):=0$. This function is nondecreasing in $x$.

At $x=\sigma_i$,
\[
\begin{aligned}
    g_{ij}(\sigma_i)
    &=
    R_j^{\mathrm{robust}}(\theta)
    \frac{\sigma_i}{B_H+e_j}\\
    &\ge
    B_{\mathrm{total}}\rho_j(\theta)
    \frac{\sigma_i}{B_{\mathrm{total}}}\\
    &=
    \rho_j(\theta)\sigma_i\\
    &\ge
    c_{ij},
\end{aligned}
\]
where we use $e_j\le B_L$ and
\[
    \rho_j(\theta)
    =
    \max_{k\in H}\frac{c_{kj}}{\sigma_k}.
\]

Thus, full commitment gives voter $i$ nonnegative payoff in market $j$. Since every positive commitment incurs the same fixed verification cost and $g_{ij}$ is nondecreasing, $\sigma_i$ is a best response in market $j$.

The participation payoff is additive across markets and voter $i$'s action set is the product box $[0,\sigma_i]^S$. Therefore,
\[
    w_i^\star=(\sigma_i)_{j\in S}
\]
is a best response in the multi-market participation game.

The argument holds for every protected voter $i\in H$ and every commitment profile of the non-protected voters. Hence both conditions of \Cref{def:settlement-liveness} hold, and the constructed reward pools provide settlement liveness for $S$.

\end{proof}

\section{Additional Simulation Results}
\label[appendix]{app:additional-simulation}

\subsection{Protected Construction Reward Cost}
\label[appendix]{app:reward-simulation-with-protected-layer}

In this section, we complement the simulation in \Cref{sec:simulation} with an evaluation of the protected-layer construction.
Unlike the preceding simulation, which treats each market as an independent single-market batch, the protected-layer construction is designed for multi-market settlement.
We therefore group our collected markets by their latest DVM round, obtaining 401 retained batches, of which 324 contain multiple markets.

\begin{table}[ht]
    \centering
    \small
    \caption{Quantiles of the required protected-layer rewards.}
    \label{tab:protected-layer-rewards}
    \begin{tabular}{lrrrr}
        \toprule
        Scenario & p50 & p90 & p95 & p99 \\
        \midrule
        Low-cost   &  \$2.53  & \$11.39  & \$21.78 & \$200.44 \\
        Baseline &  \$12.64 & \$56.95  & \$108.87 & \$1,002.19 \\
        High-cost     &  \$50.53 & \$227.80 & \$435.46 & \$4,008.73 \\
        \bottomrule
    \end{tabular}
\end{table}

\begin{figure}[htbp]
    \centering
    \includegraphics[width=0.95\linewidth]{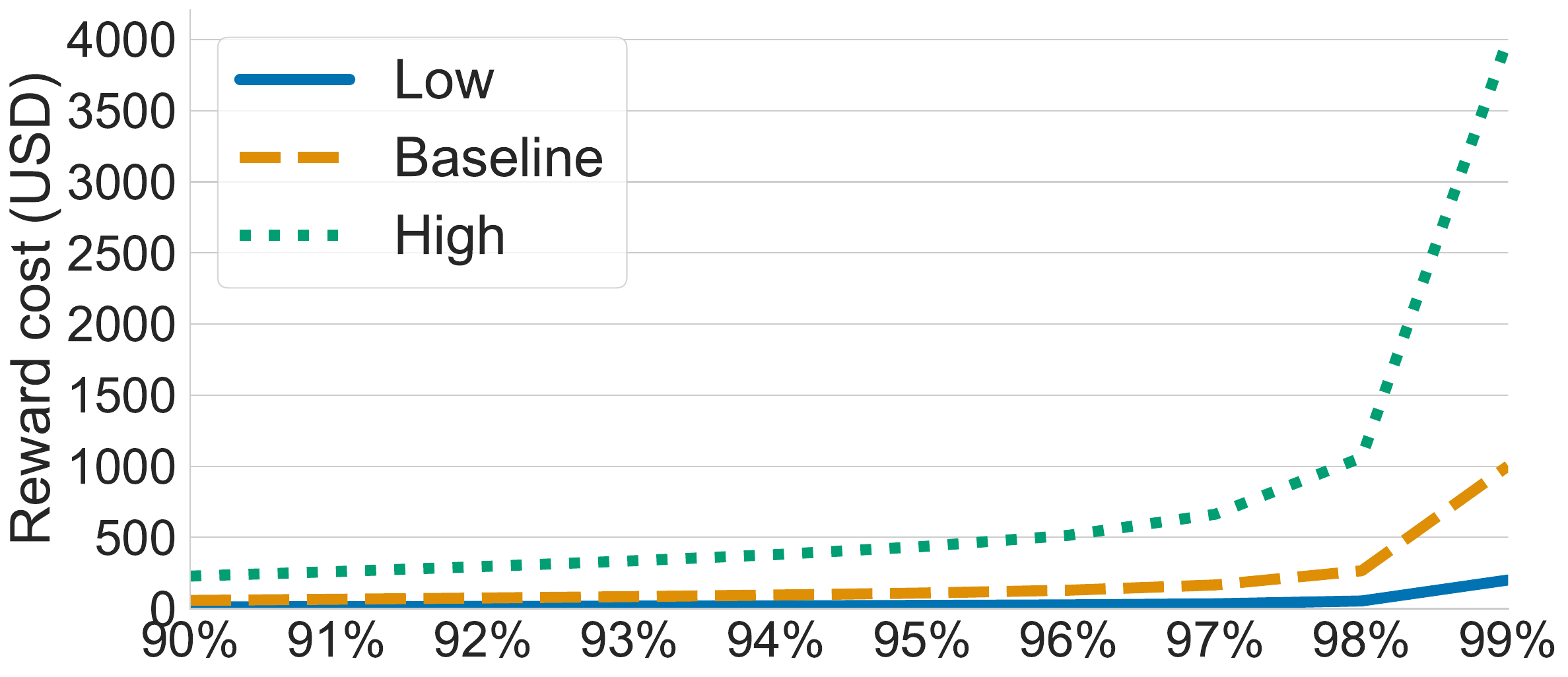}
    \caption{Required settlement rewards by the protected-layer construction under low-cost, baseline, and high-cost verification scenarios.}
    \label{fig:reward-costs-protected}
\end{figure}

We follow the same simulation procedure as in \Cref{sec:simulation}, replacing the greedy construction with our protected-layer construction. As shown in \Cref{tab:protected-layer-rewards} and \Cref{fig:reward-costs-protected}, required rewards remain moderate for the large majority of settlement batches: even under the high-cost scenario, 95\% require no more than \$500 in rewards. Substantially larger rewards arise only in the extreme tail. In practice, the protocol can treat these rare cases separately by requiring a larger bond calibrated to market exposure to cover participation rewards, or by routing settlements requiring unusually large rewards to direct second-tier resolution.

\subsection{Counterfactual Market Exposure}
\label[appendix]{app:counterfactual-market-exposure}

In this section,  we conduct a comprehensive counterfactual analysis of how the required participation reward evolves with the historical distribution of voter stake. We evaluate each historical stake snapshot across a range of hypothetical market exposures and participation-cost scenarios.

\parhead{Greedy construction}
We first evaluate the greedy construction by varying the hypothetical market open interest from \$1 million to \$5 million. For each historical voter-stake snapshot and open-interest level, we compute the required participation reward under the three participation-cost scenarios (low-cost, baseline, and high-cost). The solid line reports the median required reward across cost draws, while the shaded region shows the 5th--95th percentile range. The vertical axis is shown on a logarithmic scale.

\begin{figure}[ht]
    \centering
    \includegraphics[width=0.99\linewidth]{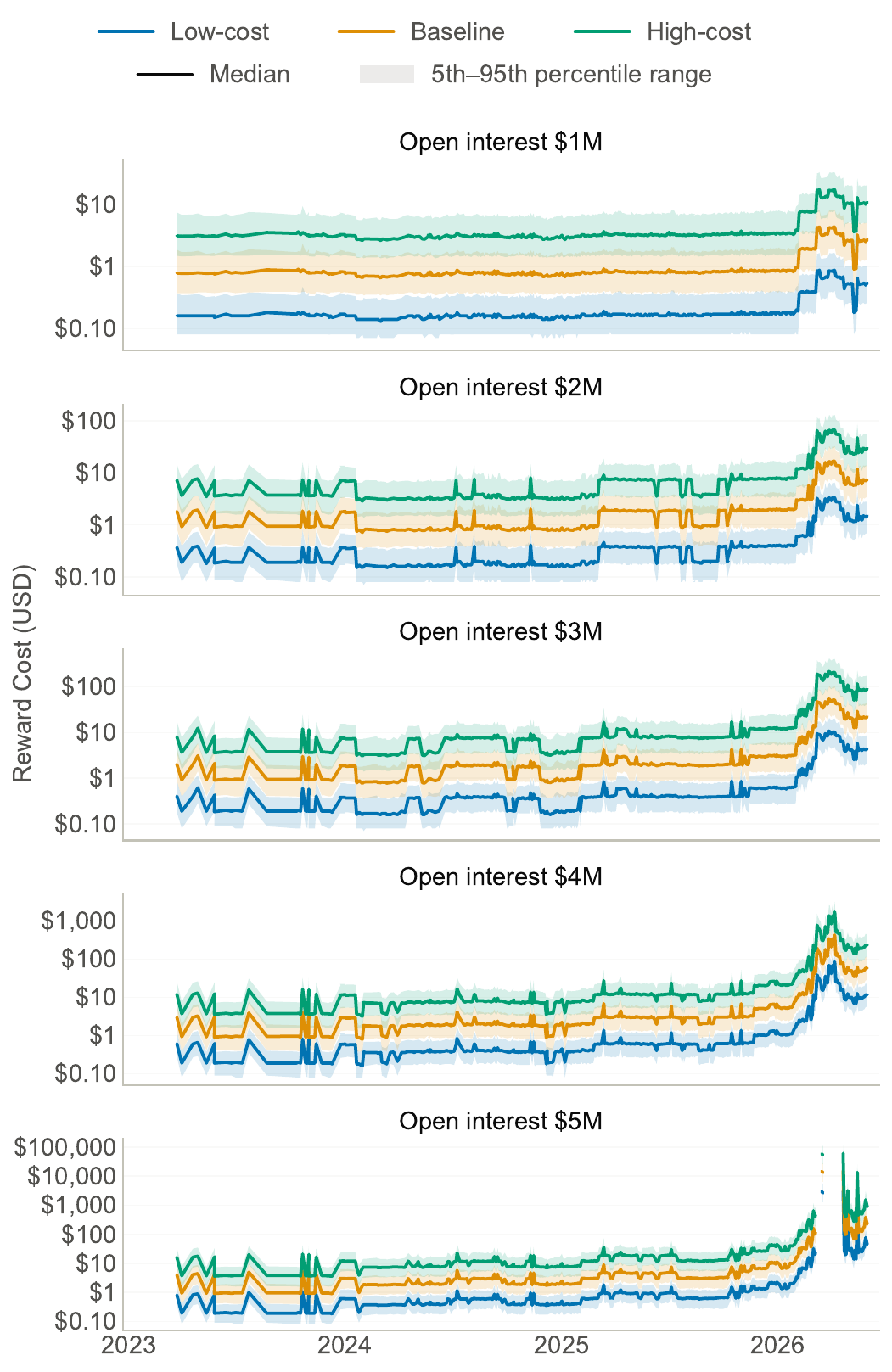}
    \caption{
    Required participation reward under historical UMA voter-stake
    snapshots and counterfactual market exposure.
    Lines show the median required reward under the low-cost ($\mu_s=\$0.20$), baseline ($\mu_s=\$1.00$), and high-cost ($\mu_s=\$4.00$) participation-cost scenarios; shaded regions show the corresponding 5th--95th percentile ranges.
    }
    \label{fig:counterfactual-reward-time-greedy}
\end{figure}

As shown in \Cref{fig:counterfactual-reward-time-greedy}, consistent with our earlier simulations, the rewards our protocol must offer voters remain low across most historical stake snapshots and counterfactual market sizes.
The required rewards increase after 2025, particularly for larger markets, as the declining dollar value of available voter stake brings a fixed market exposure closer to the settlement-capacity boundary.
Near this boundary, the greedy construction must recruit nearly all available voters, making the required reward increasingly sensitive to high participation costs.
For a \$5 million market, some historical snapshots contain insufficient available stake to satisfy the admission requirement at all; these infeasible cases appear as gaps in the figure.
More generally, very large required rewards indicate a stake-capacity bottleneck rather than a participation-cost problem. Such markets should therefore fall back to the second-tier settlement mechanism rather than rely on arbitrarily large rewards.

\parhead{Protected construction}
We further evaluate the protected construction under
counterfactual multi-market batches.
For each historical voter-stake snapshot, we fix the open interest of each market at \$1 million and vary the batch size from two to five markets, corresponding to total batch exposures from \$2 million to \$5 million.
For each snapshot and batch size, we draw market-specific participation costs under the same three cost scenarios and apply the protected construction to compute the required reward pools.

\begin{figure}[ht]
    \centering
    \includegraphics[width=0.99\linewidth]{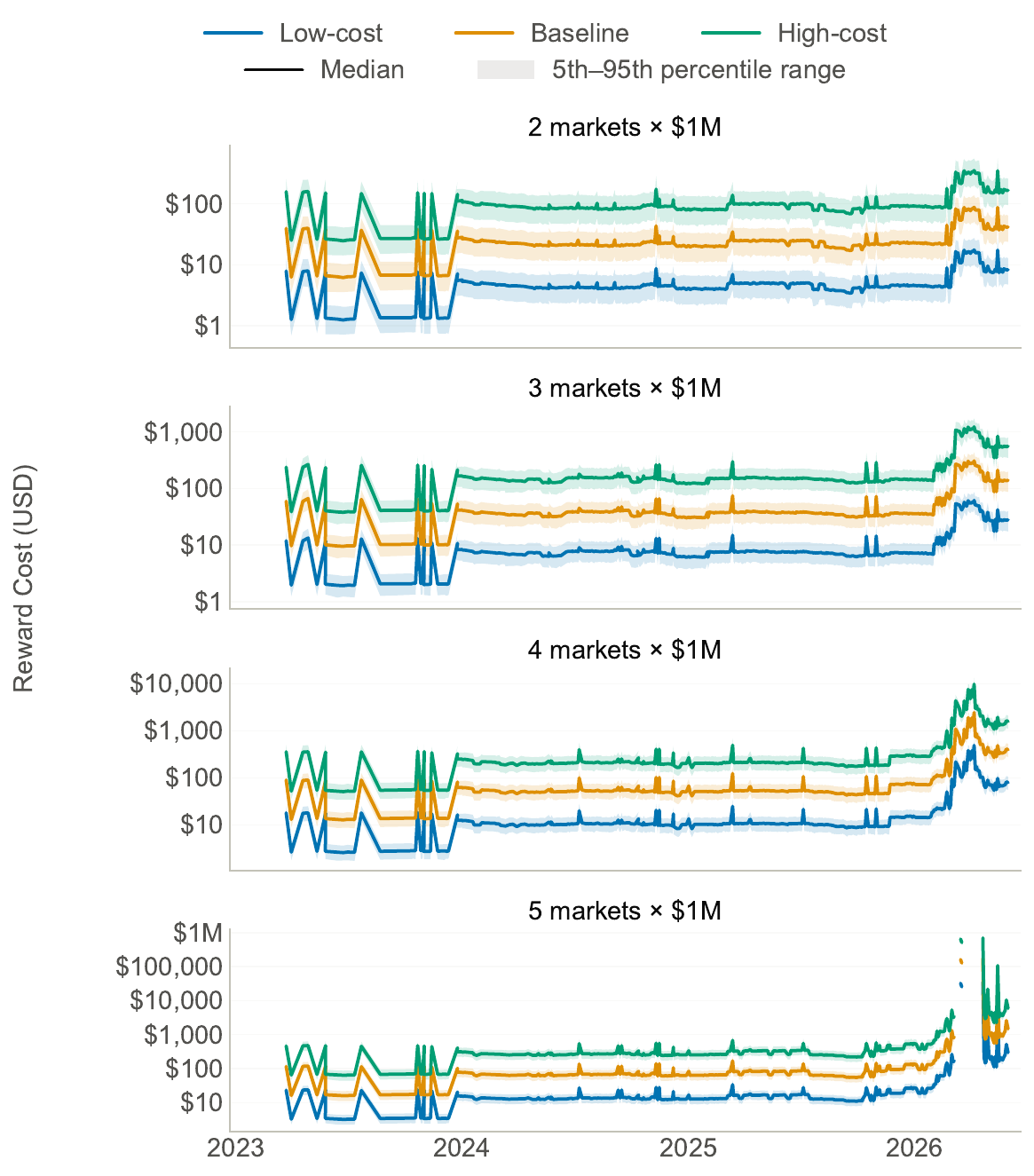}
    \caption{
    As in the single-market counterfactual analysis, the solid lines report the median required reward across cost draws, while the shaded regions show the 5th--95th percentile ranges. Gaps indicate historical snapshots for which the corresponding batch is not admission-feasible.
    }
    \label{fig:counterfactual-reward-time-protected}
\end{figure}

As shown in \Cref{fig:counterfactual-reward-time-protected}, the
protected-layer construction exhibits the same qualitative trend as the single-market greedy construction.
For batches of two to four \$1 million markets, the total rewards our protocol must offer remain modest across most historical stake snapshots, although they increase with the number of concurrent markets.
Rewards rise more sharply after 2025 as the declining dollar value of available voter stake brings larger batches closer to the settlement-capacity boundary.
For batches of five \$1 million markets, the required rewards become substantially more volatile, and some snapshots are infeasible because the available stake is insufficient to satisfy the admission
requirement. As before, these cases indicate a stake-capacity bottleneck rather than a problem that should be addressed with arbitrarily large rewards.

\end{document}